\documentclass[letterpaper,11pt]{report}

\usepackage{fullpage}
\usepackage{verbatim}
\usepackage{cite}
\usepackage{setspace}
\usepackage{fancyhdr}

\usepackage[small]{caption}
\usepackage{graphics}
\usepackage{color}

\usepackage{hyperref}
\usepackage{soul,listings,xcolor}
\usepackage{url}

\usepackage[dvips]{graphicx}

\def\title{Mytitle}

\definecolor{dkgreen}{rgb}{0.2,0.6,0}
\definecolor{gray}{rgb}{0.5,0.5,0.5}
\definecolor{mauve}{rgb}{0.58,0,0.82}

\begin{document}

\def\addrone{Your address}
\def\addrtwo{Your city}

\def\degree{M.Tech. in Computer Science with Specialization in Information Security}

\def\submissiondate{January 08, 2011}

\def\supervisorone{Dr. Ponnurangam Kumaraguru}

\def\supervisortwo{Dr. Vinayak Naik}

\def\supervisorthree{Dr. Sundeep Oberoi}

\thispagestyle{empty}

\begin{center}

{\LARGE \bf {Broker Bots: Analyzing automated activity during High Impact Events on Twitter}

 }  
 \vspace{1in}
 
 {\Large{Student Name: Sudip Mittal}} \\  
 \vspace{.1in} 
 IIIT-D-MTech-CS-IS-11-001 \\

 Nov 30, 2014 \\
  
    \vspace{.6in}

  \vspace{.5in}

{Indraprastha Institute of Information Technology\\
New Delhi}

\vspace{.8in}  {\underline{Thesis Committee}} \\ \supervisortwo                 \\ \supervisorthree \\ \supervisorone ~(Chair) \\ \vspace{.35in}

\vspace{.6in}
 {Submitted in partial fulfillment of the requirements \\for the Degree of M.Tech. in Computer Science}

\vspace{.8in}

\copyright 2014 Sudip Mittal \\ All rights reserved \\
\vspace{.8in}

\end{center}


\newpage

\pagestyle{empty}
\vspace*{7.1in} 

\newpage

\begin{center}
\section*{Certificate}\label{section:certificate}
\end{center}
This is to certify that the thesis titled \textbf{``Broker Bots: Analyzing automated activity during high impact events on Twitter"} submitted by \textbf{Sudip Mittal} for the partial fulfillment of the requirements for the degree of \emph{Master of Technology} in \emph{Computer Science \& Engineering} is a record of the bonafide work carried out by her / him under my / our guidance and supervision at the Cybersecurity Education and Research Centre, Indraprastha Institute of Information Technology, Delhi (CERC@IIITD). This work has not been submitted anywhere else for the reward of any other degree. \\ \vspace{0.5in}

\textbf{Professor PK}\\
\textbf{Indraprastha Institute of Information Technology, New Delhi}

\begin{abstract}

Twitter is now an established and a widely popular news medium. Be it normal banter or a discussion on high impact events like Boston marathon blasts, February 2014 US Icestorm, etc., people use Twitter to get updates and also broadcast their thoughts and views. Twitter bots have today become very common and acceptable. People are using them to get updates about emergencies like natural disasters, terrorist strikes, etc., users also use them for getting updates about different places and events, both local and global. Twitter bots provide these users a means to perform certain tasks on Twitter that are both simple and structurally repetitive, at a much higher rate than what would be possible for a human alone. During high impact events these Twitter bots tend to provide a time critical and a comprehensive information source with information aggregated form various different sources.

In this study, we present how bots participate in discussions and augment them during high impact events. We identify bots in 5 high impact events for 2013: Boston blasts, February 2014 US Icestorm, Washington Navy Yard Shooting, Oklahoma tornado, and Cyclone Phailin. We identify bots among top tweeters by getting all such accounts manually annotated. We then study their activity and present many important insights. We determine the impact bots have on information diffusion during these events and how they tend to aggregate and broker information from various sources to different users. We also analyzed their tweets, list down important differentiating features between bots and non bots (normal or human accounts) during high impact events. We also show how bots are slowly moving away from traditional API based posts towards web automation platforms like IFTTT, dlvr.it, etc. Using standard machine learning, we proposed a methodology to identify bots/non bots in real time during high impact events. This study also looks into how the bot scenario has changed by comparing data from high impact events from 2013 with data from similar type of events from 2011. Bots active in high impact events generally don't spread malicious content. Lastly, we also go through an in-depth analysis of Twitter bots who were active during 2013 Boston Marathon Blast. We show how bots because of they're programming structure don't pick up rumors easily during these events and even if they do; they do it after a long time. 

\end{abstract}

\newpage
\pagestyle{empty}

\newpage

\section*{Acknowledgments}\label{section:acknowledgments}
\pagestyle{plain}
\pagenumbering{roman}

I thank all members of Precog research group and CERC@IIITD (Cybersecurity Education and Research Centre) at IIIT-Delhi for their valuable feedback and suggestions. I would also like to thank Aditi Gupta and Paridhi Jain for their feedback and support. Last and not the least, I really appreciate and thank ``PK" for being an awesome advisor and mentor.

\newpage

\tableofcontents
\listoffigures 
\listoftables

\newpage

\newpage

\newpage
\mbox{}


\chapter{Introduction Research Aim and Contribution}\label{chapter:introduction}
\pagenumbering{arabic}
\setcounter{page}{1}
\onehalfspacing
\section{Introduction}
Twitter has been transformed into a news and media source, it is being used by all top news agencies like CNN\footnote{\url{https://twitter.com/cnnbrk}}, BBC\footnote{\url{https://twitter.com/BBCBreaking}}, etc., and also by print media like New York Times\footnote{\url{https://twitter.com/NYTLive}}, The Huffington Post\footnote{\url{https://twitter.com/HuffingtonPost}}, etc., all vying to redirect huge numbers of Twitter users to their websites and content. These sources tend to push news to Twitter.

Now, Twitter is even being used by people like politicians, celebrities to create news on Twitter too, Twitter is being used by US Senators and Congress representatives to update the media and their constituents. On twitter its even possible for normal users to create news; Abbottabad resident Sohaib Athar created a huge sensation when he live tweeted the US raid to capture Osama Bin Laden.\footnote{\url{http://techcrunch.com/2011/05/02/heres-the-guy-who-unwittingly-live-tweeted-the-raid-on-bin-laden-2/}} Twitter is a medium to detect popularity; politicians, celebrities are even paying a lot to get more and more followers.\footnote{\url{http://www.dailymail.co.uk/news/article-2430875/Barack-Obama-19-5m-fake-Twitter-followers.html}}

Our work focusses on Twitter because of its immense impact on news and news content generation. Also, data collection on Twitter is easier because of its well maintained and mature APIs. 

\section{Defining High Impact Events}
Everyday there are scores of events that can be categorized as news worthy. Events like Government policy changes, elections, earthquakes, celebrity gossip, etc., all can be categorized as news. 

\textbf{What makes an event news-worthy?} An event is news-worthy if it has: Timing, significance, Proximity, Prominence, Human Interest.\footnote{\url{http://www.mediacollege.com/journalism/news/newsworthy.html}}

We however only include high impact events in our study. We define \textbf{``high impact events"} as those events that have a great political and economical impact; they may also have high to moderate damages to life and property. 

\section{Use of Twitter in High Impact Events}
During high impact events there is a huge increase in activity on Twitter. Users all across the world login to check for news, discuss, express their sympathies, opinions, and share content. 

\subsection{Use of Twitter during the Boston Marathon Blast (April 15, 2013)}
In our dataset of tweets regarding Boston Blasts we encountered many tweets:
\begin{itemize}
\item I saw people's legs blown off. Horrific. Two explosions. Runners were coming in and saw unspeakable horror.  \textit{-- Jackie Bruno (@JackieBrunoNECN) April 15, 2013}
\item At the ER. Not a comforting way to pass the time.\#boston So sad. \textit{-- GanderHeroDog (@veterantraveler) April 15, 2013}
\item Prayers goes out to those involved/hurt in \#BostonMarathon. WTF is wrong with people man. Just sad \textit{-- LeBron James (@KingJames) April 15, 2013}
\item An eyewitness during the explosions at the \#Boston \#Marathon says ``it sounded like a cannon blast." Video: on.cnn.com/1399A40 \textit{-- CNN Video (@CNNVideo) April 15, 2013}
\item If you are concerned about a runner in the \#BostonMarathon, you can see where they last checked in here theh.gr/XCP0II \textit{-- taylor johnson (@thehappygirl) April 15, 2013}
\end{itemize}

\section{Use of Bots Online}
Internet bot, web robot, WWW robot or simply bot are software application that runs automated tasks over the Internet. Bots perform tasks that are simple and repetitive, at a higher rate than would be possible for a human. Bots can be good or have malicious intent. Malicious use of bots include orchestrating DDoS attacks, spreading spam campaigns, commit click fraud, etc. Examples of Internet Bots: Wikipedia bots\footnote{\url{http://www.bbc.com/news/magazine-18892510}}, Twitter bots, Spam bots, IRC bots, Botnets, etc.

\section{Use of bots on Twitter (\textit{Twitter bots})}
Twitter bots are Twitter users that post updates to Twitter automatically. They are used many a times to help in spam campaigns, directing Twitter users to certain webpages, post directed messages, and sometimes to assist Twitter users by updating them about information like highway traffic, natural disaster alerts, etc. 

Some examples of Twitter bots:
\begin{itemize}
\item \textit{@AllOilPainting} : Helps user search for oil paintings being sold on Ebay. Tweets the links to the painting for its followers.

\item \textit{@Horse\_ebooks} : Horse\_ebooks is a spam bot that is followed by approximately 200,000 users. Its famous for its amusing  ``non sequiturs" updates. Horse\_ebooks was named one of the best Twitter feeds, by UGO Networks in 2011\footnote{\url{http://www.ugo.com/web-culture/best-twitter-accounts-of-2011-horse-ebooks}} and Time.com in 2012.\footnote{\url{http://techland.time.com/2012/03/21/the-140-best-twitter-feeds-of-2012/slide/horse-ebooks/}} John Hermann at Splitsider wrote that Horse\_ebooks ``might be the best Twitter account that has ever existed".\footnote{\url{http://splitsider.com/2012/01/the-ballad-of-horse_ebooks/}}

\item \textit{@earthquakeBot} : It's an emergency bot that tweets about earthquakes. 
\begin{figure}[!ht]
\centering
\includegraphics[scale=.4]{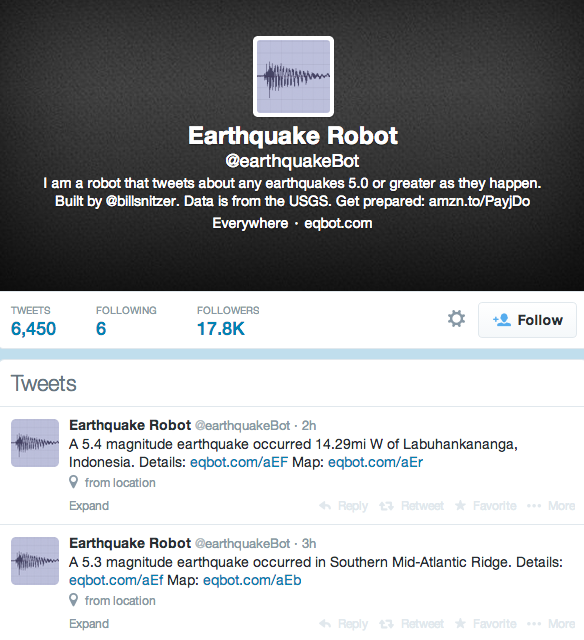}
\caption [Example of an Automated Account.]{Tweets from the \textit{@earthquakeBot}.}
\label{fig:earthqk}
\end{figure}
\end{itemize}
\section{Rules for Posting Updates on Twitter}\label{rules}
An understanding of the Twitter API rules for ``posting updates" to Twitter is vital in designing Twitter bots, because if an account flouts any of these rules they are either banned from posting updates to Twitter for some time or their account is suspended.

Formal documentation by twitter is available for these ``Twitter rules"\footnote{\url{https://support.twitter.com/articles/18311-the-twitter-rules}} and ``API Limits".\footnote{\url{https://support.twitter.com/articles/15364-twitter-limits-api-updates-and-following}}

Some major rules: 
\begin{enumerate}
\item Update limits: 
\begin{itemize}
\item Direct messages: 250 per day.
\item Tweets: 1,000 via API per day. The daily update limit is further broken down into smaller limits for semi-hourly intervals. Retweets are counted as tweets.
\item Following (daily): The technical follow limit is 1,000 per day.\footnote{\url{https://support.twitter.com/articles/68916-following-rules-and-best-practices}}
\item Following (account-based): Once an account is following 2,000 users, additional follow attempts depend upon account-specific ratios. 
\end{itemize}
\item Content boundaries and Use of Twitter:
\begin{itemize}
\item Don't follow and/or unfollow large amounts of users in a short time period, particularly by automated means.
\item Don't repeatedly follow and unfollow people, whether to build followers or to garner more attention for your profile.
\item Updates should not consist mainly of links.
\item Don't post duplicate content over multiple accounts or multiple duplicate updates on one account.
\item Don't post multiple unrelated updates to a topic using \#, trending or popular topic, or promoted trend.
\item Don't send large numbers of duplicate @replies or mentions.
\item Don't send large numbers of unsolicited @replies or mentions in an aggressive attempt to bring attention to a service or link.
\item Don't randomly or aggressively Retweet through automation in an attempt to bring attention to an account, service or link.
\item Don't randomly or aggressively Favoriting tweets through automation in an attempt to bring attention to an account, service or link.
\end{itemize} 
\end{enumerate}

\section{Research Motivation}
As bots get more and more popular on Twitter they will tend to impact and affect discussions, information flow, credibility, information security, etc. During high impact events people rely on twitter for important updates, crucial information. We were motivated to analyze how bot activity affects high impact events. 

\section{Research Contribution}\label{chapter:contri}
We in this section list down all the major contributions of this study. 

\begin{itemize}
\item We show that bots actively participating in high impact events spread information obtained from ``trusted and verified" sources. 
\item We show that bots aid in information distribution, they help in ``brokering" information from various sources to other users. In the Boston marathon blasts we show that bots push updates to at least 9.53\% new users. 
\item We show that bots do not propagate rumors and even if they do; they do it after some time. 
\item Bots are moving away from Twitter API based approach to web automation softwares like IFTTT, dlvr.it in order to post updates to Twitter. 
\item We show that temporal beaded Twitter features do not actually add value in differentiating bot and non-bot accounts on Twitter. 
\item We create a classifier based on user based features with an accuracy of 85.10\% in classifying bot and non-bot accounts. 
\end{itemize}

\section{Computer Science Contribution}
The work done in this document falls under the panel ``Computational Social Science". Computational social science includes study of computing systems and social interactions. We in our work analyze behavior of bots during high impact events on Twitter. We approach this problem from a more computer science perspective. We begin by discussing our elaborate data collection and annotation schemes. We then proceed to compare different features for bots and non-bot accounts one by one to create a method to differentiate between the two classes. We provide insights into how bots function during high impact events. 

\vspace{1.5cm}
The rest of the study is structured in the following manner. In Chapter 2 we cover the related work. Chapter 3 covers how we selected 5 high impact events followed by our data collection, annotation, and enrichment methodology. Our Analysis is covered in Chapter 4. Chapter 5, covers discussion, limitations and future work. 

\chapter{Related Work}\label{chapter:related_work}
This chapter presents previous work done in understanding High impact events and automated activity on Twitter. We first discuss work done on high impact events and then work on Twitter bots.
\section{Work on High Impact Newsworthy Events}
A large number of studies have been conducted to understand high impact events on Twitter. 
Kwak et. al. \cite{Kwak:2010:TSN:1772690.1772751} discussed whether Twitter is a social network or a news media. They showed that nearly 85\% topics talked about on Twitter were in fact related to news. 
Palen et. al. \cite{Palen:2008:EOW:1460563.1460583} analyzed Twitter adoption and use during emergencies. 
Zhao et. al. \cite{Zhao:2011:CTT:1996889.1996934} 222. 
Mendoza et. al. \cite{Mendoza:2010:TUC:1964858.1964869} analyzed the use of Twitter for emergency response activity during the 2010 earthquake in Chile.
Castillo et. al. \cite{Castillo:2011:ICT:1963405.1963500} worked on ``Information Credibility" on Twitter. They showed that automated classifiers can be used to find information and determine its credibility. 
Gupta et. al. \cite{Gupta:2012:CRT:2185354.2185356} created a mechanism to rank credible information on Twitter. 
Longueville et. al. \cite{DeLongueville:2009:OHI:1629890.1629907} used location based mining to get spatio-temporal data on forest fires in France. 
Sakaki et. al. \cite{Sakaki:2010:EST:1772690.1772777} used Twitter to locate epicenter and impact of earthquakes. 
Agrawal et. al. \cite{Oh:2011:ICT:1968924.1968971} tracked the Mumbai terrorist attack on Twitter. 
Verma et. al. \cite{VermaVCPMPSA11} used NLP techniques to extract ``situational awareness tweets" during emergencies.
Vieweg et. al. \cite{Vieweg:2010:MDT:1753326.1753486} analyzed the use of Twitter during two natural hazards events.
Gupta et. al. \cite{gupta20131} analyzed fake content on Twitter during the Boston marathon blast. 

\section{Work on Automated Activity (\textit{bots}) on Twitter} \label{RWB}

Zhang et. al. \cite{Zhang:2011:DAA:1987510.1987521} worked on analyzing time trends present in bot activity on Twitter. The argued that in order to avoid being banned from posting tweets to Twitter, bots have to ensure that they do not exceed the 1,000 tweets per day Twitter API limit. So they must post approximately 41 tweets per hour or approximately 1 tweet every 0.68 minute, this will create a pattern in their tweeting behavior. 
Chu et. al. \cite{Chu:2010} They obtained about 6,000 accounts annotated to be either human, bot or cyborg (bot account that is ``curated" by a human) accounts. They studied various set of features that can help distinguish between these three type of accounts.
Roure et. al. \cite{de2013observing} observed ``social machines in the wild". They monitored interactions with bots and bot lifecycle on Twitter. They argued that the ``purpose" of the bot is the key attribute that must be studied for analyzing bots. 
Boshmaf et. al. \cite{boshmaf2011socialbot} proved that online social networks are vulnerable to infiltration by bots on a large scale. They showed that it is very easy to run astroturf campaigns to spread misinformation and propaganda.
Messias et. al. \cite{messias2013you} studied social bots and their influence over the network in which they were active. They said that popular influence scores are vulnerable and can be easily manipulated. 
Tavares et. al. \cite{tavares2013scaling} did many temporal analysis to compare activity of bots and humans. 
Wald et. al. \cite{Wald}  studied the users that were susceptible to bots on Twitter. They approached it from the view point of spam bots and argued that user influence score was a key factor in determining if that user will engage a bot. 

\section{Types of Bots discussed in literature.}
In Section \ref{RWB} we discussed major work done on bots on Twitter. Different types of bots have been discussed in literature. In this Section we will discuss various types of bots discussed in papers mentioned in Section \ref{RWB}:
\begin{enumerate}
\item \textbf{Explicit Bots}: The class of bots who declare that they are bots. This can be done through mentioning the same in their profile description. Many a times these bots also mention their creators by mentioning their twitter handles. 
\item \textbf{Implicit Bots}: These are the bots that do not mention that they are bots. 
\item \textbf{Social Bots}: Bots which interact with other users on Twitter. 
\item \textbf{Retweet Bots}: Bots who only Re-Tweet tweets by other users. These bots can Re-Tweet based on particular keywords or only Re-Tweet tweets by some particular user. Example of these type of bots are those accounts who Re-Tweet all popular tweets about say ``cricket' or ``Boston"; another example can be accounts which Re-Tweet all tweets by popular users like  ``@katyperry" or ``@BillGates".
\item \textbf{News Bots}: These are a special class of Retweet Bots which Retweet tweets with content from important news sources like ``@cnnbrk". They also Retweet content based on certain keywords associated with the type of news being propagated by these accounts. The need for a separate class arrises because these are the accounts which begin heavy retweeting when a high impact event occurs. 
\item \textbf{Cyborgs}: The type of a Twitter account that is ``curated" by a human. 
\end{enumerate}

\section{Difference from previous work}
Both Chu et. al. \cite{Chu:2010} and Zhang et. al. \cite{Zhang:2011:DAA:1987510.1987521} have in their work highlighted methods to distinguish automated activity from human activity on Twitter. They presented innovative schemes to distinguish between the two classes during the normal everyday functioning of these accounts. We on the other hand present insights and interesting findings on bot accounts that participate in discussion during high impact events. We also show that methods proposed by Chu et. al. \cite{Chu:2010} and Zhang et. al. \cite{Zhang:2011:DAA:1987510.1987521} do not provide good results when applied to bots actively participating high impact events. We in our work suggest improvements to better differentiate and understand automated activity during high impact events.

\chapter{Methodology}\label{chapter:method}

In this Section we are going to discuss how we selected the events that were analyzed, followed by our data collection, annotation, and enrichment methodology.

\section{Event Selection}

We selected 5 high impact events from the year 2013 and 2014 for analysis. The events selected included natural hazards, political and terror strikes. The main theme associated with these events was that they had huge impact worldwide and gained traction all across the world. These events in our opinion caused a huge economic, political waves throughout the world along with loss of lives and property. We refer to these events as ``high impact events." 
		 	 	 		

The table \ref{table:events} describes the data collection details of 5 events we selected, it also includes all the hashtags associated with each selected event.

Event description:
\begin{enumerate}
\item Boston blasts: On April 15, 2013, 2:49 pm EDT 2 bombs exploded during the Boston Marathon 2013. 3 people died and 264 were injured in the bombing. Bombings were followed by shooting, manhunt and firefight. 

\item Oklahoma Tornado: 2013 Moore tornado struck Moore, Oklahoma, and adjacent areas on the afternoon of May 20, 2013, killing 24 people damaging about \$2 billion worth of property in Grady, McClain, and Cleveland counties in Oklahoma; particularly the city of Moore.

\item Washington Navy Yard Shooting: lone gunman Aaron Alexis on September 16, 2013, shot 12 people and injured 3 others at the Naval Sea Systems Command (NAVSEA) inside  Washington Navy Yard. The attack began around 8:20 a.m. EDT and ended around 9:20 a.m. EDT; when Aaron Alexis was killed by police.

\item Cyclone Phailin: Very Severe Cyclonic Storm Phailin stuck Thailand, Myanmar, India on October 4, 2013, till October 14, 2013. Causing damages of about \$696 million (USD 2013) and 45 fatalities. In India the cyclone wreaked havoc in Anadaman and Nicobar Islands, Andhra Pradesh, Orissa, and Jharkhand. 

\item Ice-storm: During the period of February 11 to 17, 2014; a fierce ice-storm struck the East and the South coast of USA There were about 22 fatalities and damages of about 15 million USD.

\end{enumerate}

\begin{table}[!h]
\begin{center}
    \begin{tabular}{|l|l|l|}
    \hline
    Name               & Hashtags & Data Collection Duration\\ \hline
    Boston Blasts      & Dzhokhar Tsarnaev, \#Watertown, & April 15 - April 21, 2013 \\& \#manhunt, Sean Collier, & \\&\#BostonStrong,& \\& \#bostonbombing, \#oneboston, & \\& bostonmarathon,& \\& \#prayforboston, & \\& boston marathon, \#bostonblasts, & \\& boston blasts, bostonterrorist,& \\& boston explosions, bostonhelp, & \\& boston suspect &  ~     \\ \hline
        Oklahoma Tornado   & Oklahoma, tornado,&May 20 - May 22, 2013  \\& PrayForOklahoma, & \\& care4kidsOK        &          \\ \hline
        Navy Yard Shooting & NavyYardShooting,& September 16 - September 18, 2013 \\& navy yard shooting,& \\& Washington navy yard         &          \\ \hline

    Cyclone Phailin     & phailin, cyclonephailin        &    October 4 - October 16, 2013       \\ \hline
    Ice-storm           & Icestorm,  \#USIcestorm, &February 11 - February 19, 2014 \\& \#SnowInUS        &          \\ \hline
   
    \end{tabular}
\caption[Events in consideration]{Details about various selected events.}
\label{table:events}
\end{center}
\end{table}


\section{Event Data Collection}

We used the Twitter Streaming API to collect data about the events. 

\begin{table} [!h]
\tiny
\begin{center}
    \begin{tabular}{|l|l|l|l|l|l|}
    \hline
    Metric&Boston Blast Marathon & Icestorm & Navy Yard Shooting & Oklahoma Tornado & Cyclone Phailin \\ \hline
    Total tweets & 7,888,374 & 433,880 & 484,609 & 809,154 & 76,136 \\ \hline
    Total unique users & 3,677,531 & 198,391 & 257,682 & 542,049 & 34,776\\ \hline
    Tweets with URLs & 3,420,228 & 233,576 & 290,887 & 388,541 & 44,990 \\ \hline
    Number of re-tweets & 5,217,769 & 209,556 & 262,362 & 509,732 & 41,718\\ \hline
    Start date & April 15, 2013 & February 11, 2014 & September 16, 2013& May 20, 2013 & October 4, 2013 \\ \hline
      End date & April 21, 2013 & February 19, 2014 & September 18, 2013& May 22, 2013 & October 16, 2013 \\ \hline
    \end{tabular}
\caption[Data collected]{Details about collected data.}
\label{table:eventsdet}
\end{center}
\end{table}

Statistics and numbers associated with the dataset for each High Impact Event: (Table \ref{table:eventsdet})


\section{Data Annotation}

After collecting the data we compiled a list of top 200 users for each of the 5 events based on the number of tweets they posted in context to the event. This group of 1000 high frequency users were then manually annotated by a group of Masters and PhD students in Computer Science at IIIT Delhi. The criteria for selecting these annotators was that they must be familiar with Twitter landscape and must have used Twitter for a minimum period of 1 year. These set of annotators were given the following instructions: 
\\

\begin{lstlisting}
What is a bot?: Twitter bots are, essentially, computer programs that tweet on their own accord. While normal people access Twitter through its Web site and other clients, bots connect directly to the Twitter via its API (application programming interface), parse information and post to Twitter automatically.

There are various kinds of bots like retweet bots, news bots, humor bots

Example of a bot to tweet about earthquakes in San Francisco:

https://twitter.com/earthquakesSF 

User description: I am a robot that tells you about earthquakes in the San Francisco Bay area as they happen. I get my data right from the USGS. I was programmed by @billsnitzer.

This account specifies that it is a bot.
A user may or may not do so.
If a user does not mention use the TIPS mentioned below.

What is to be done here?
STEP 1: Visit the URL of the Twitter profile.
STEP 2: Study the account. (Have a look at the TIPS)
STEP 3: Mark whether an account is a BOT account or not.

TIPS: While making the decision pay attention to the following:
Closely read the user description.
Most bots have a large number of tweets.
Time difference between tweets: Bots tweet very frequently
Number of Friends and Followers: Most Bots have many friends and few followers
Bots usually tweet about some specific topic or Re-Tweet tweets from particular users.

If you come across accounts that you think are being ``curated" by humans. We request you not to mark them as bots. 

Decide if the account is a Bot or not. 
Mark the choice. 
Wherever you are not sure mark: Can't Decide.
\end{lstlisting}

Every account in the set of the above mentioned 1000 users was annotated by 3 annotators.

Each annotator was given a list of account URLs and 3 choices per account:
\begin{itemize}
\item Bot.
\item NOT a Bot.
\item Can't Decide. 
\end{itemize}

We did not consider ``Cyborgs" as bots in our analysis. We took a very strict approach in labeling the annotated accounts as Bots and Non-Bots. All those accounts that were annotated as Bots by all 3 annotators were labelled as Bots. A similar strict bound was applied to the accounts labelled as Non-Bots. We used this highly strict criteria for labeling so as to ensure very good quality data. At the end of this stage we had 2 distinct classes of accounts; namely Bots and Non-Bots.  

\begin{table}[!h]
\begin{center}
    \begin{tabular}{|l|l|l|}
    \hline
    Event              & Bots & Non-Bots \\ \hline
    Boston Blasts      & 97    & 17       \\ \hline
    Icestorm           & 87    & 7        \\ \hline
    Navy Yard Shooting & 90    & 11       \\ \hline
    Oklahoma Tornado   & 47    & 42       \\ \hline
    Cyclone Phailin     & 56    & 38        \\ \hline
    Total (Includes accounts active in multiple events)     & 377    & 115        \\ \hline 
    \end{tabular}
\caption[Annotation result]{Number of accounts in each class after annotation.}
\end{center}
\end{table}

\section{Data Enrichment}
After creating a true positive data-set for both bot and non bot categories for 5 high impact events, we proceeded to get more data about these Twitter accounts to enrich this data for further analysis. 
Using various Twitter API endpoints we collected profile meta-data and all of their friends and followers. We also collected maximum possible tweets for all of the above mentioned annotated accounts.

\chapter{Analysis}\label{chapter:analysis}

In this section, we start with analyzing some basic characteristics of bots and then move on to the analysis of their followers. We also discuss their network, impact on information diffusion and rumor propagation during high impact events. We present an in-depth analysis on URLs, Tweet source and Tweet time. We then discuss all the important features that can help us predict if an account is a bot or not during high impact events. We try to predict once using all user based and temporal features and once after discarding temporal features and compare the results of the two. We also present a detailed analysis of a few bots that were active during different high impact events. We also compare bot activity from 2013 to bot activity in 2011. 

\section{Exploratory Numbers and Bot Creation Methodology}\label{createbot}
Our annotated data consists of 377 bots that were active during 5 different events out of which 26 bots were active in multiple events. We have 309 distinct annotated bots and 115 distinct annotated non-bots. We look at 5 events from 2013 and 7 events from 2011. In order to successfully study Twitter bots first we must understand the logic and how they are created. In order to avoid being banned from posting tweets to Twitter, bots have to ensure that they do not exceed the 1,000 tweets per day Twitter API limit. So they must post approximately 41 tweets per hour or approximately 1 tweet every 0.68 minute. They must also follow all rules discussed in Section \ref{rules}.

There are 4 major ways to create bots on Twitter:
\begin{enumerate}
\item Popular Tweet based bots: These bots can ``listen" for tweets that are currently popular on Twitter via searching for popular tweets (flags present in the API) or they can post content from ``Twitter Trending topics". They repost (Retweet or copy and post content as their own) these tweets.
\item Keyword based bots: These bots look for certain keywords on Twitter and repost tweets that contain them.
\item Source based bots: These bots repost content from specific Twitter users. For example, a bot can repost all updates from news accounts like ``@cnnbrk", ``@BBCWorld", etc., these type of bots are termed ``News Aggregators".\footnote{\url{http://www.zillman.us/white-papers/bots-blogs-and-news-aggregators/}}
\item Outside content based bots: These bots look for updates outside Twitter like RSS and other web feeds, updates to blogs, dedicated databases , etc., and post these updates to Twitter usually with a link to the same. 
\end{enumerate}

\section{Bots active During Multiple Events}
In our dataset we found that 26 bots were active during multiple events. We manually analyzed the ``Twitter Timeline" of all 26 of these bots and found that 8 were Keyword based bots, 9 were source based bots, 2 were Popular Tweet based bots and 7 were outside content based. 

\section{Characteristics}
\subsection{Followers and Friends on Twitter}
\begin{figure}[!ht]
\centering
\includegraphics[scale=.4]{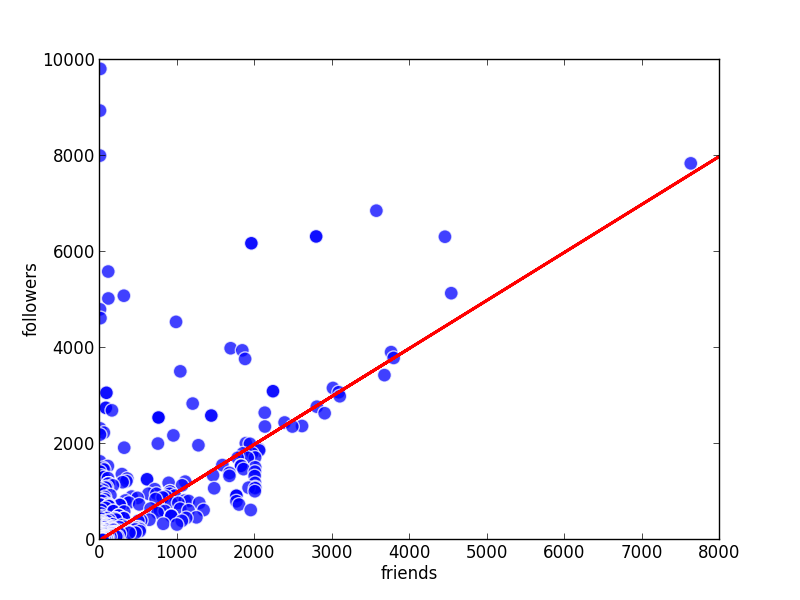}
\caption{Followers vs Friends ratio for bots.}
\label{fig:ff1}
\end{figure}
\begin{figure}[!ht]
\centering
\includegraphics[scale=.4]{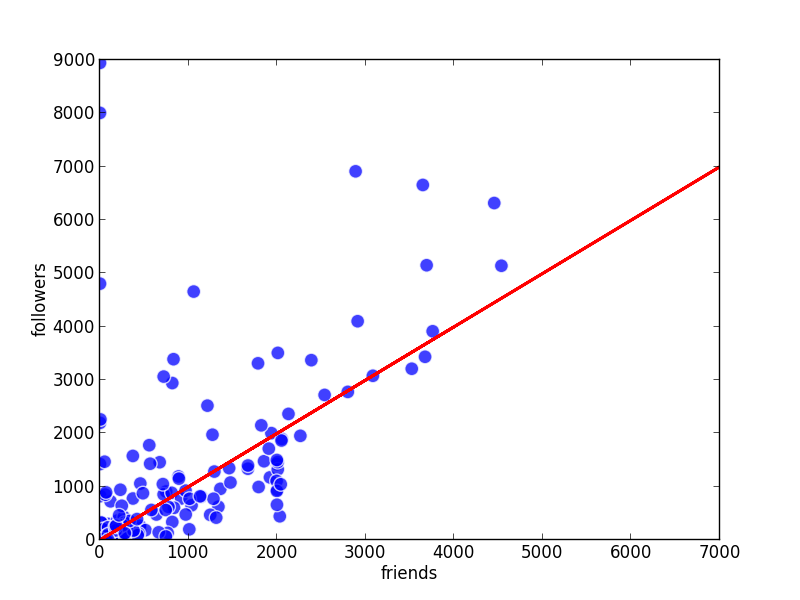}
\caption{Followers vs Friends ratio for non-bots.}
\label{fig:ff2}
\end{figure}

We created graphs for Followers and Friends (Figure \ref{fig:ff1} and \ref{fig:ff2}) for both bots and non-bots. The graph is more spread out for non-bots. Data points for bots were more clustered around the origin. Similar results were obtained by Chu et. al. \cite{Chu:2010}.

\subsection{Profile Description on Twitter}

\begin{figure*}[!ht]
\centering
{\includegraphics[scale=.25]{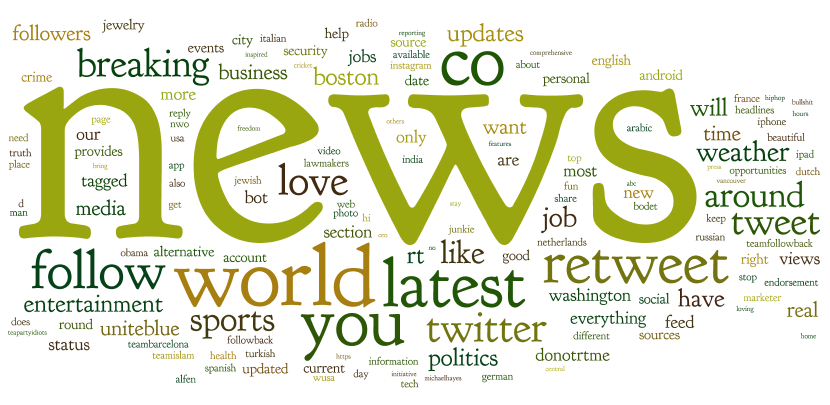}}
\label{fig:wordle1}
{\includegraphics[scale=.275]{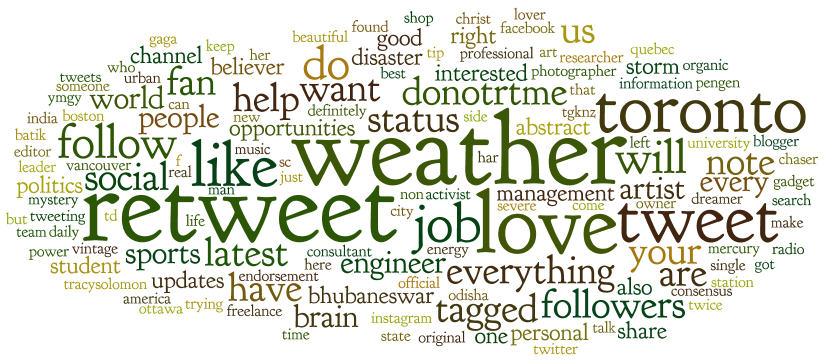}}
\label{fig:wordle2}
\caption[Frequent words present in the profile description of bots and nonbots.]{Most frequently used words in profile description for bots (left) and non-bots(right).  Size of the word denotes frequency.}\label{fig:wordle}
\end{figure*}

We found out top words used by both bot and non-bot accounts in their profile description on Twitter (Figure \ref{fig:wordle}). The most frequent words used by bots highlight the motive behind creating these bots that are active during high impact events which can be inferred from words like ``news", ``world", ``breaking", ``retweet", ``latest", etc. Whereas on the other side one can find more general words that are used to describe non-bot accounts. 


\section{Bot Friends and Data Source Analysis}\label{srcv}

We wanted to find out the most prominent / common sources from where bots pick their data during high impact events. One way we can measure it can be through a count of all the ``@mentions" which is used on Twitter to hold conversations or cite the source of a particular Tweet. An example for the same can be seen in Figure \ref{fig:cred}, here the bot is citing @GP\_Today which is the Twitter account of the website gptoday.com that publishes Formula 1 news articles. The bot account in question listens to its web feed updates and tweets them giving reference to the source via ``@mentions" to improve the credibility of the news \cite{Castillo:2011:ICT:1963405.1963500},\cite{Gupta:2012:CRT:2185354.2185356}. 

\begin{figure}[!h]
\centering
\includegraphics[scale=.6]{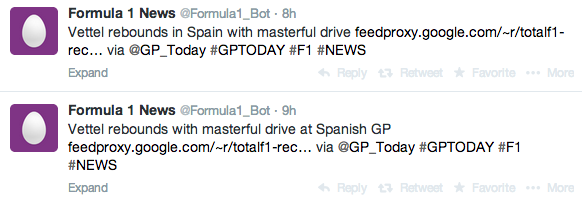}
\caption[Example of a bot giving credit.]{A news bot giving credit to its source using the ``via" keyword. This bot is picking up content from a web feed and posting it on Twitter.}
\label{fig:cred}
\end{figure}

In case of Retweet bots, they keep on listening for updates from other Twitter accounts and Retweet them as soon as new updates are posted. Retweets texts in the Twitter API results appear as ``\textit{RT @mention Tweet text...}". 

\begin{figure}[!h]
\centering
\includegraphics[scale=.6]{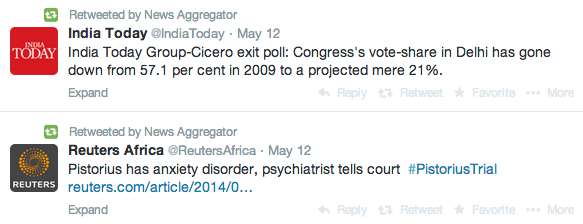}
\caption{A Retweet bot reposting content form credible news sources.}
\label{fig:cred}
\end{figure}

Hence creating just a frequency table of @mention will give us the top sources from where bots pick their data during high impact events. Users mentioned by accounts that mention other users with whom they are interacting (high frequency directed or targeted mentions) will not actually impact the top 15 data sources during high impact events as the frequency of such @mentions will be low.

\begin{figure}[!h]
\centering
\includegraphics[scale=1]{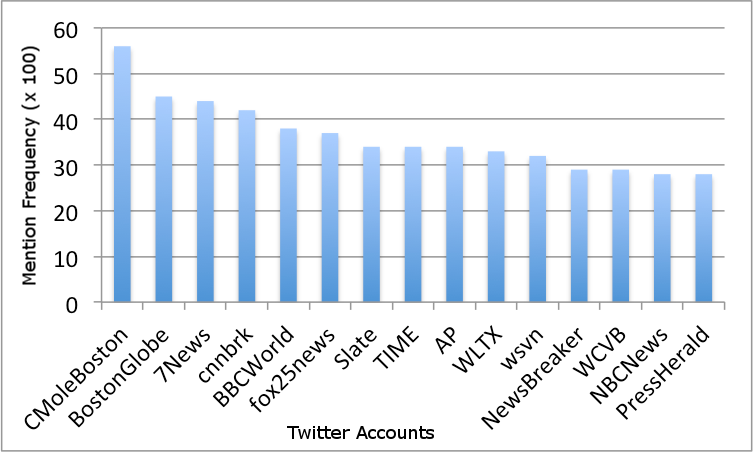}
\caption{Top sources during Boston blasts.}
\label{fig:topsour}
\end{figure}

Figure \ref{fig:topsour} shows the top 15 data sources and their frequency in the Boston marathon blasts. Out of all the 4,562 accounts mentioned by annotated bots active during the Boston marathon blasts 652 (14.29\%) were actually verified. Table \ref{table:veri} lists the top 15 data sources and if they were verified accounts or not. 

\begin{table}[!h]
\centering
    \begin{tabular}{|l|l|}
    \hline
    \textbf{Source} & \textbf{Verified/Non-verified} \\ \hline
CMoleBoston & Non-verified \\ \hline
    BostonGlobe  & Verified     \\ \hline
    7News        & Verified     \\ \hline
    cnnbrk       & Verified     \\ \hline
    BBCWorld     & Verified     \\ \hline
    fox25news    & Verified     \\ \hline
    Slate        & Verified     \\ \hline
    TIME         & Verified     \\ \hline
    AP           & Verified     \\ \hline
    WLTX         & Verified     \\ \hline
    wsvn         & Verified     \\ \hline
    NewsBreaker  & Non-verified \\ \hline
    WCVB         & Verified     \\ \hline
    NBCNews      & Verified     \\ \hline
    PressHerald  & Non-verified \\ \hline
    \end{tabular}
    \caption[Top data sources and if they are verified ]{Top 15 data sources and if they are verified by Twitter or not.}
    \label{table:veri}
\end{table}

\section{Bot Followers}

In this section we will analyze the users who follow bots that are active during high impact events. We discuss the bot network and their impact on Information diffusion during high Impact events. For our analysis we collected all followers of bots using Twitter API endpoint. The total number of unique bot followers collected were 623,198. 

\subsection{Profile Description of the Followers.}

\begin{figure}[!h]
\centering
\includegraphics[scale=.4]{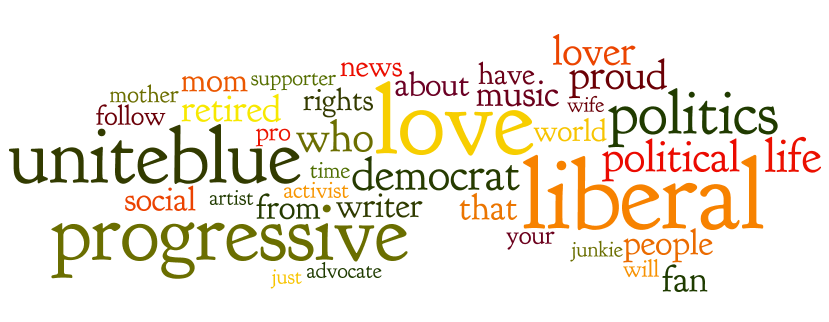}
\caption[Common words in profile description of bot followers.]{Most frequent words in profile description of bot followers in the Boston bast dataset. Size of the word denotes frequency.}
\label{fig:follwe}
\end{figure}

In Figure \ref{fig:follwe} we present the top frequency words common in the profile description of these bot followers.

\subsection{Bot and Follower Network in Boston Marathon Blasts}
To study the bot and their follower network, we collected bot followers of all the 97 annotated bots in our dataset. The total number of unique bot followers collected were 623,198.

\begin{figure}[!hbtp]
\centering
\includegraphics[scale=.7]{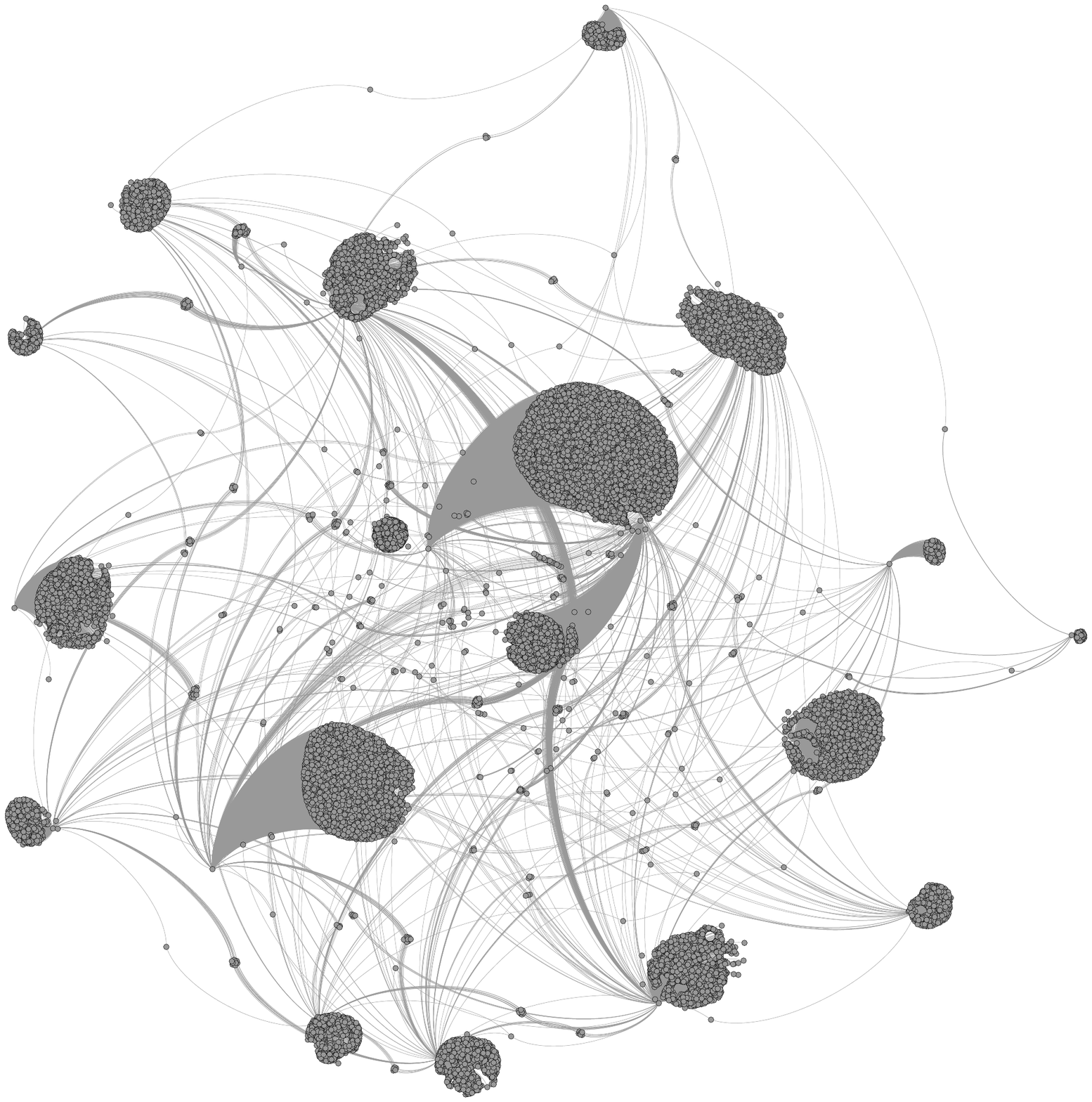}
\caption[Network graph of 40,000 randomly selected bot followers.]{Network graph of 40,000 randomly selected bot followers. Average degree is 1.01545}
\label{fig:network4}
\end{figure}

\begin{figure}[!hbtp]
\centering
\includegraphics[scale=.65]{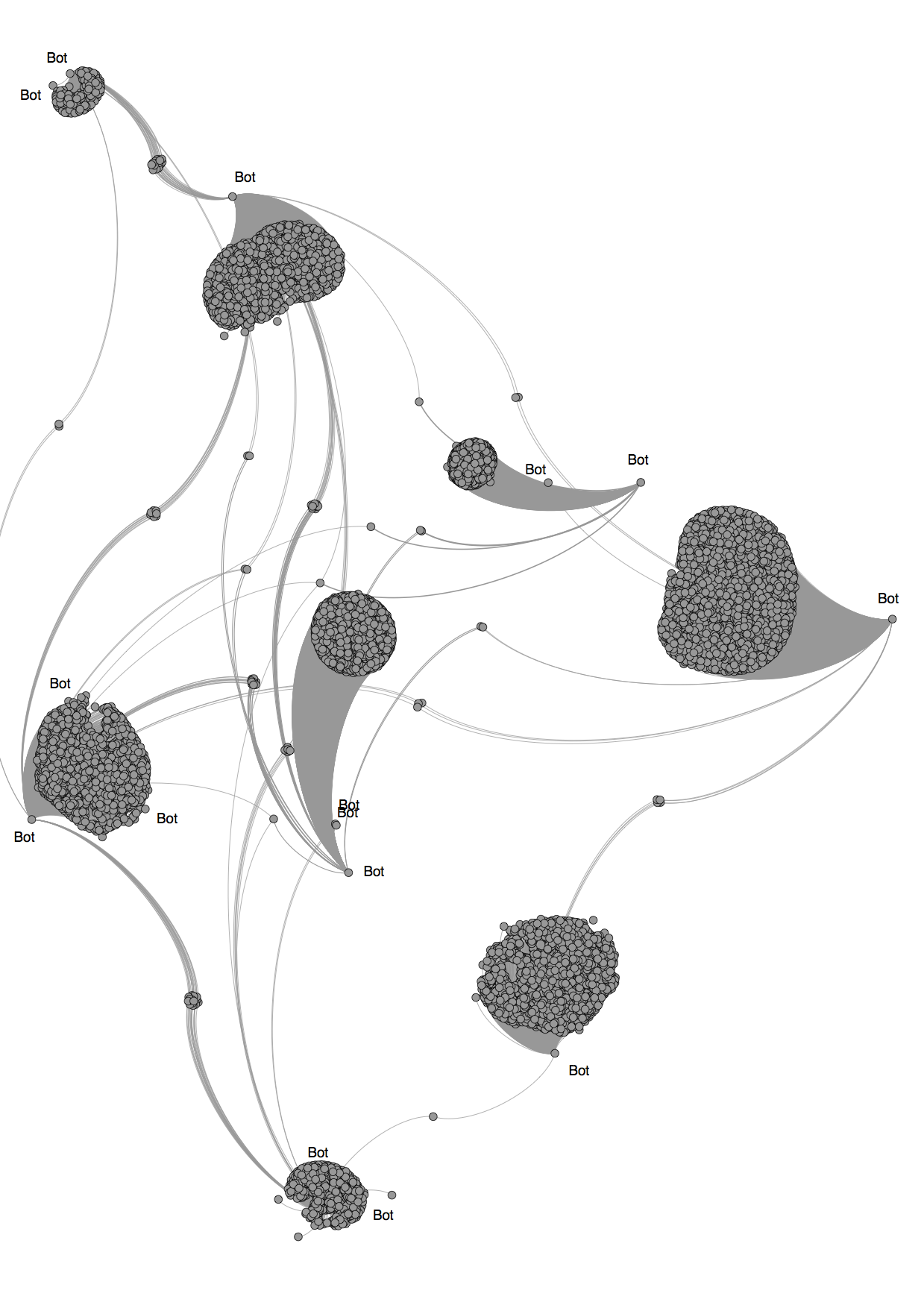}
\caption{Network graph of 20,000 randomly selected bot followers}
\label{fig:network2}
\end{figure}

We created network graph for a random set of 40,000 followers (Figure \ref{fig:network4}) and also a graph for 20,000 randomly selected followers (Figure \ref{fig:network2}).\footnote{We could not create the graph for all users because of memory restrictions.} Both these graphs were created in Gephi - an open graph viz platform.\footnote{\url{https://gephi.org/}} Though we wanted to create and analyze graphs of all 623,198 bot followers and 97 bots from the Boston marathon blast dataset, we were limited by memory restrictions.  For Figure \ref{fig:network4} the average degree for the graph was 1.01545. 

Looking at Figure \ref{fig:network2} and Figure \ref{fig:network4} and that the average degree for Figure \ref{fig:network4} is 1.01545. We can safely conclude that many users tend to follow very few bot accounts. Most of these accounts follow only one bot account in our Boston blast high impact event, very few accounts follow more than one bot account. 

We can attribute this behavior to the fact that users would not like to follow many high frequency bot accounts who Tweet about the same thing as these accounts who Tweet a lot will flood a users Twitter Timeline with their content. 

\newpage
\subsection{Impact on Information Diffusion during High Impact Events.}

\begin{figure}[!h]
\centering
\includegraphics[scale=.35]{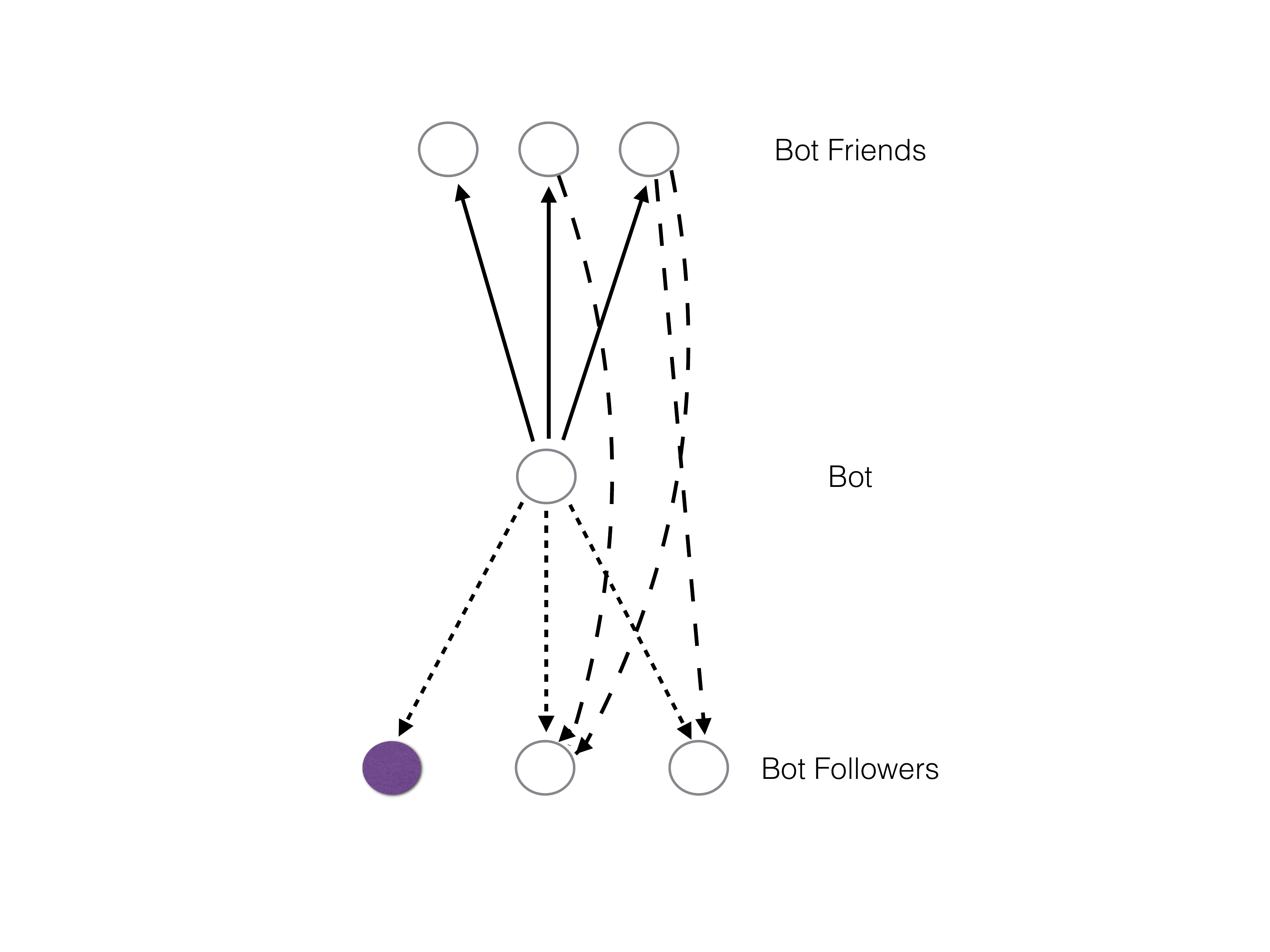}
\caption[Study of impact due to automated accounts on information diffusion during high impact events]{A simple node diagram showing relation between, bots, bot friends and followers. A follower can follow bot friends too. The colored node represents the bot followers who don't follow bot friends and hence, get updates about high impact events from bots.}
\label{fig:inflow}
\end{figure}

To answer the question, how many new users received information as a result of automated activity on Twitter during high impact events and what was the impact of bots on information diffusion. We focussed our study on the Boston blast dataset. We collected all bot followers of the 97 annotated bot accounts found in the Boston marathon dataset. We then collected their friends. We used this methodology in place of collecting the followers of bot friends because a very popular bot friend was ``@cnnbrk" which at the time of writing this document had close to 17 million followers and collecting details of these 17 million followers is nearly impossible and there are many more bot followers. 

Using all the above mentioned data we then created a list of all accounts following bots and all accounts among these accounts that followed bot friends. We then found out all accounts that get updates only from bots and not from bot friends (represented by colored node in Figure \ref{fig:inflow}). In our dataset for Boston blast marathon these accounts were 9.53\% of all bot followers. This analysis suggest that \textit{at least 9.53\% of bot followers are getting fresh updates from these bots during high impact events as a result of them following bots and not accounts that are the most likely ``verified" source of this information} (see table \ref{table:veri}). However, posts by these bots are also visible in various search results and the public Twitter Timeline further increasing the reach of information to many different users. We can also loosely suggest that bots aid information diffusion during high impact events.  These bots can be seen as ``\textbf{brokering}" information to other users.

\subsection{Role in Rumor Propagation}\label{rumor}
During Events there is lot of conflicting information that is spreading on online social networks. Castillo et. al \cite{Castillo:2011:ICT:1963405.1963500}.  and Gupta et. al. \cite{Gupta:2012:CRT:2185354.2185356}.  in their respective papers have highlighted the spread of misinformation on Twitter and have also suggested methods to find credible information on Twitter. Similar analysis was done by Gupta et. al. in their paper \cite{gupta20131}, discussed the spread of fake and malicious content on Twitter during the Boston Marathon bombings. We will focus our analysis on spread of rumors during the same Boston marathon bombings by bots. 

Some key rumors that were propagated regarding the bombings on Twitter were:\footnote{\url{http://www.snopes.com/politics/conspiracy/boston.asp}}
\begin{enumerate}
\item Suspect became citizen on 9/11: One rumor suggested that the suspects became naturalized citizens of the US on September 11.  Tweet text: ``\textit{RT @pspoole: Dzhokhar Tsarnaev received US citizenship on Sept 11, 2012 -- 11 years to the day after 9/11 attacks http://t.co/kHLL7mkjnn}". The 1st tweet was by @pspoole on Friday April 19, 2013, at 15:34:56 (+0000); this rumor was shared 2,321 times by unique users in our dataset.
\item Sandy Hook Child Killed in Bombing: One rumor claimed that one victims was an 8-year-old girl who attended Sandy Hook school and was running for the victims of the Sandy Hook shootings. Tweet text: ``\textit{RT @CommonGrandma: She ran for the Sandy Hook children and was 8 years old. \#prayforboston http://t.co/cLir6nI7tB}". 1st tweet on Friday, April 19, 2013, at 09:56:45 (+0000); this rumor was shared 1,743 times by unique users in our dataset.
\item Donating 1\$ Tweet:\footnote{\url{http://globalnews.ca/news/482442/fake-boston-marathon-donation-twitter-account-suspended/}} A Tweet by a fake account @\_BostonMarathon, ``\textit{For each RT this gets, \$1 will be donated to the victims of the Boston Marathon Explosions. \#DonateToBoston}" went viral on Twitter. Tweet time: Monday, April 15, 2013, at 11:29:23 (+0000); this rumor was shared 28,350 times by unique users in our dataset.

\end{enumerate}

Findings from our dataset concerning bots: 

\begin{enumerate}
\item Suspect became citizen on 9/11: In our dataset only 2 bots Retweet this rumor on Friday, April 19, 2013, at 15:41:35 and 15:47:58 (+0000).
\item Sandy Hook Child Killed in Bombing: This rumor was never picked up by any of the 97 bots in our dataset.
\item Donating 1\$ Tweet: Only 1 bot in our dataset picked up this Tweet that too on Wednesday, April 17, 2013 at 00:50:24 (+0000) 
Gupta et. al. claim that the Tweet was Retweeted 28,350 times in their paper \cite{gupta20131}, while working on the \textit{same} dataset. 
\end{enumerate}

This particular bot behavior of not picking up and spreading rumors can be attributed to the fact that they get their data from verified sources (see Section \ref{srcv} and table \ref{table:veri}). 

\section{Analysis of Tweets}
In this section, we analyzed tweets posted by bots during high impact event. We will first put forward our analysis of URLs present in these tweets, then move on to discuss our findings regarding the source of the Tweet. We will also present our results from time related analysis. 

\subsection{URL Analysis}\label{URLA}
In our Boston Blast data we have 44,071 tweets by accounts annotated as bots and 7,099 tweets by annotated non-bots. Out of these tweets 36,672 (83\%) bots tweets and 4,849 (68\%) non-bots tweets had URLs. Table \ref{table:url} lists the top URL hostnames shared during Boston Marathon Blasts by frequency.\footnote{Similar results were observed in the other 4 High Impact events.} Top 8 hostnames used by Boston Bots are standard URL shorteners out of which bit.ly is popular among both bots and non-bots (Site cmole.com has been suspended, at the time of writing this document). To further analyze we expanded all the bit.ly shortened URLs to get their expanded URLs. The top 10 bit.ly shortened hostnames have been listed in the Table \ref{table:urlbitly}.

\begin{table}[!h]
\begin{center}
    \begin{tabular}{|l|l|l|}
    \hline
    Rank & Boston Bots     & Boston Non Bots    \\ \hline
    1    & bit.ly          & satu-Indonesia.com \\ \hline
    2    & j.mp            & bit.ly             \\ \hline
    3    & dlvr.it         & adf.ly             \\ \hline
    4    & q.gs            & inktothepeople.com \\ \hline
    5    & cmole.com      & tavernkeepers.com  \\ \hline
    6    & adf.ly          & twitter.com        \\ \hline
    7    & bo.st           & www.youtube.com    \\ \hline
    8    & youtu.be        & apne.ws            \\ \hline
    9    & pulpnews.com    & nbcnews.to         \\ \hline
    10   & www.rightnow.io & cnsnews.com        \\ \hline
    \end{tabular}
\caption{Top URL hostnames shared during Boston Blast.}
\label{table:url}
\end{center}
\end{table}

\begin{table}[!h]
\begin{center}
    \begin{tabular}{|l|l|l|}
    \hline
    Rank & Boston Bots          & Boston Non Bots      \\ \hline
    1    & www.sigalert.com     & feeds.abcnews.com    \\ \hline
    2    & www.indeed.com       & news.google.com      \\ \hline
    3    & likes.com            & rss.cnn.com          \\ \hline
    4    & boston.craiglist.org & feedproxy.google.com \\ \hline
    5    & www.youtube.com      & edition.cnn.com      \\ \hline
    6    & feeds.abcnews.com    & aol.sportingnews.com \\ \hline
    7    & news24s.com          & www.theblaze.com     \\ \hline
    8    & feedproxy.google.com & www.news12.com       \\ \hline
    9    & declassifieds.info   & www.mediaite.com     \\ \hline
    10   & www.woweather.com    & www.guardian.co.uk   \\ \hline
    \end{tabular}
\caption{Top bit.ly expansion URL hostnames shared during Boston Blast.}
\label{table:urlbitly}
\end{center}
\end{table}

To further check if these URLs were malicious or not, we compared them with google's database using the Google Safe Browsing API.\footnote{\url{https://developers.google.com/safe-browsing/}} Out of the 36,672 bot posted URLs and 4,849 non-bots posted URLs we found that only 188 URLs posted by bots and 27 posted by non-bots were marked as ``Malicious". As the percentage of malicious URLs is very low, we can say that bots participating in discussions on high impact events generally do not spread malicious content. 

\subsection{Tweet Source Analysis} \label{TSA}
The Twitter API gives us the source from where the Tweet originated for example Twitter for iPhone, web, etc. This field provides an excellent feature to differentiate bot and non-bot accounts. Table \ref{table:bostonsource} list top sources as observed in Boston Marathon Blast event.\footnote{Similar results observed for other events. See Appendix.} 

\subsubsection{Use of Existing Automation Software}
In the top Tweet source for bots one can see that many automation services are being used. Services like ``dlvr.it" and ``IFTTT" are being used heavily to direct content from outside sources onto Twitter. ``dlvr.it" is an online service that automatically publishes an RSS feed to various social media like Twitter, Facebook, Google+ etc., ``IFTTT" lets user create simple ``if then else" statement that are triggered via outside Twitter activity like a new blog, a new Instagram picture etc., and the action is implemented/published on Twitter or any other service mentioned by the user.\footnote{See Appendix for some IFTTT recipes.} The use of these services makes it very easy to create bots on Twitter. A bot creator does not even need to have programming experience or dedicated machines to run bots, they can user the service provided by these web-apps. 

A possible reason for using automated systems can be to circumvent coding and bot hosting expenditures. These automation services provides simple web based GUI's through which bot creators can add bot logic to post content onto Twitter as mentioned in Section \ref{createbot}. All these automation services require from the bot creators is to log into Twitter once in order to create app API keys; using which these services are able to post updates on Twitter.

\begin{table}[!h]
\begin{center}
    \begin{tabular}{|l|l|l|l|l|}
    \hline
    Rank & Boston Bots Source              & Count & Boston Non Bots Source & Count \\ \hline
    1    & twitterfeed                     & 11405            & Web                    & 2603             \\ \hline
    2    & web                             & 5844             & twitterfeed            & 1969             \\ \hline
    3    & Tweet Old Post                  & 4052             & Tweet Button           & 1042             \\ \hline
    4    & dlvr.it                         & 3962             & Twitter for iPhone     & 453              \\ \hline
    5    & IFTTT                           & 2049             & TweetDeck              & 298              \\ \hline
    6    & TweetDeck                       & 1515             & Botize                 & 246              \\ \hline
    7    & Crime News Updates              & 950              & Twitter for iPad       & 220              \\ \hline
    8    & VenturaCounty\_Retweets         & 609              & Echofon                & 216              \\ \hline
    9    & WordPress.com                   & 530              & Twitterfall            & 31               \\ \hline
    10   & Strictly Tweetbot for Wordpress & 353              & Instagram              & 12               \\ \hline
    11   & Bitly Composer                  & 186              & HootSuite              & 3                \\ \hline
    \end{tabular}
\caption{Top Tweet Sources for Boston Blast dataset.}
\label{table:bostonsource}
\end{center}
\end{table}

\subsection{Tweet Time Analysis} \label{TTA}
Zhang et. al. in their paper \cite{Zhang:2011:DAA:1987510.1987521}, based their hypothesis for detecting ``Automated Activity" on Twitter on the fact that highly automated accounts will exhibit timing patterns that are not found in non-automated users. They argued that ``Automated activity is invoked by job schedulers that execute tasks at specific times or intervals". They assumed that bots in order to maximize reach while keeping in mind the constraint of posting only 1,000 tweets in a day need to spread their tweets to at least 1 Tweet per minute in order to escape penalty under Twitter Limits.

 \begin{figure*}[h]
\centering
{\includegraphics[scale=.35]{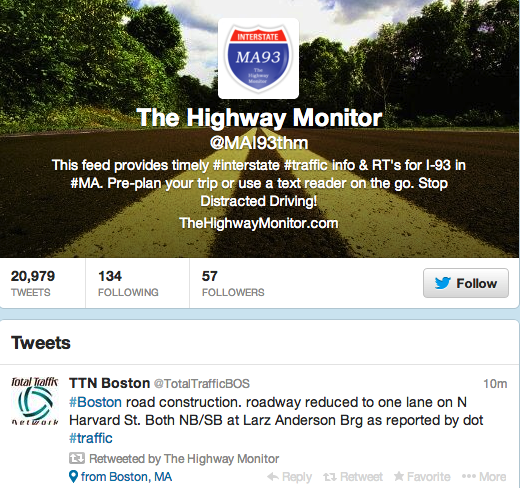}}
\label{fig:subfig2}
{\includegraphics[scale=.35]{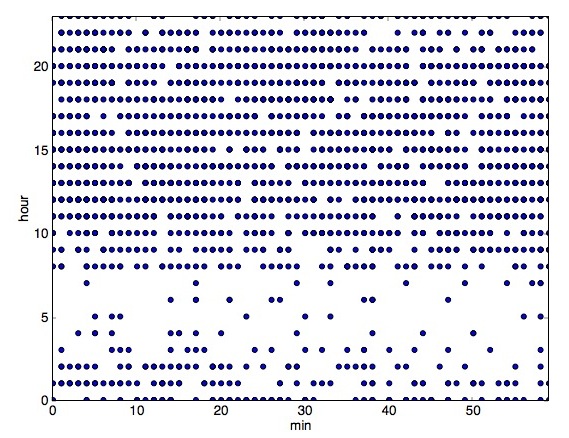}}
\label{fig:subfig3}
\caption[Time analysis of a bot active during Boston blasts]{A bot from Boston dataset, and its corresponding tweeting activity based on its Timeline.}
\label{fig:fake}
\end{figure*}

\begin{figure}[!h]
\centering
\includegraphics[scale=.35]{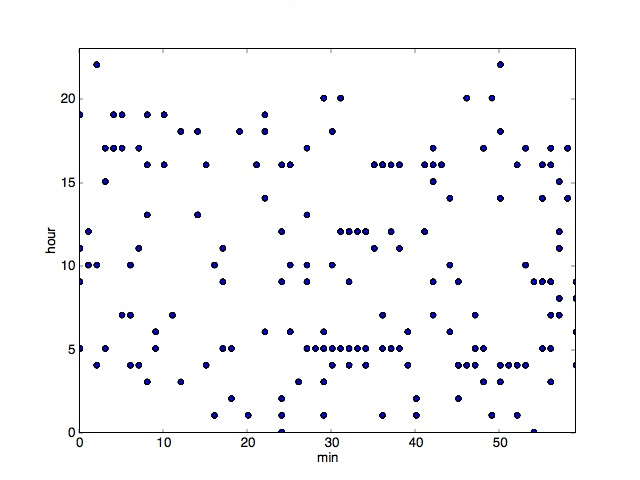}
\caption[Time analysis of a nonbot active during Boston blasts]{A non-bot from Boston dataset, and its tweeting activity based on its Timeline.}
\label{fig:nonbottweet}
\end{figure}

We observed similar results as observed by Zhang et.al. (See Figure \ref{fig:fake} and Figure \ref{fig:nonbottweet}). Both these plots were based on their entire Twitter Timeline for a week. Usually most of the ``high impact events" tend to last only for a few days sometimes even hours. During this short period of time it becomes really tough to differentiate between a bot and a non-bot account using data only from the event. Both high frequency bot and non-bot accounts show similar tweeting pattern (see Figure  \ref{fig:fake}) 

 \begin{figure*}[!h]
\centering
{\includegraphics[scale=.35]{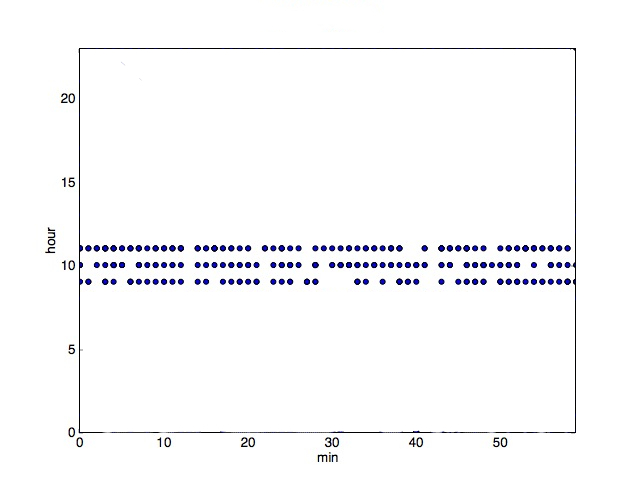}}
\label{fig:subfig2}
{\includegraphics[scale=.35]{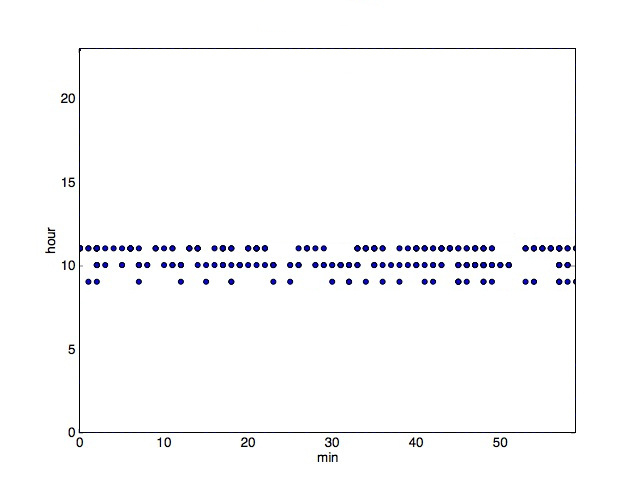}}
\label{fig:subfig3}
\caption[Tweeting pattern of a bot and a nonbot during Boston blasts]{Similar Tweeting pattern observed for a bot account (left) and a non-bot account (right) in the Boston blast event.}
\label{fig:fake2}
\end{figure*}

\begin{figure*}[!h]
\centering
{\includegraphics[scale=.45]{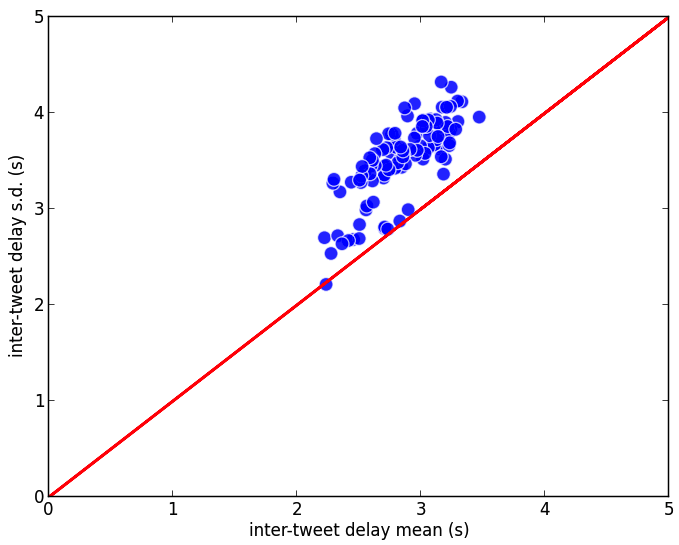}}
\label{fig:subfig2}
{\includegraphics[scale=.45]{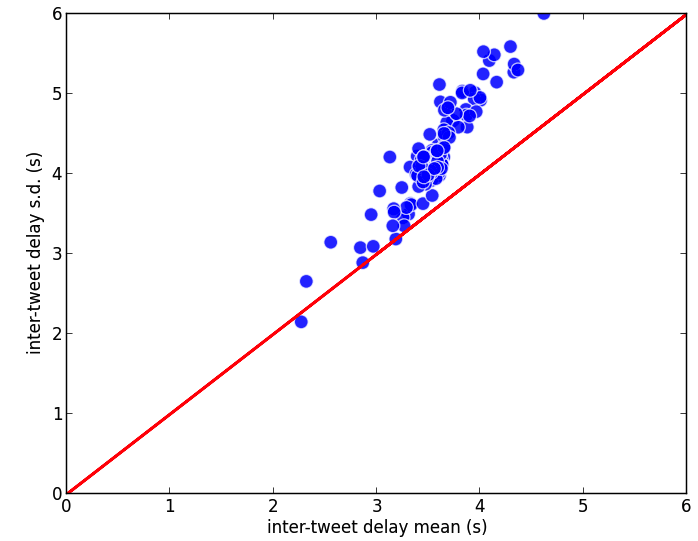}}
\label{fig:subfig3}
\caption[Inter tweet time mean and standard deviation for bot and non-bot accounts]{Inter tweet time mean and standard deviation for bot accounts (left) and non-bot accounts (right) using their data only from the events. (All bots and all non-bots from all events).}
\label{fig:fake3}
\end{figure*}

\begin{figure*}[!h]
\centering
{\includegraphics[scale=.28]{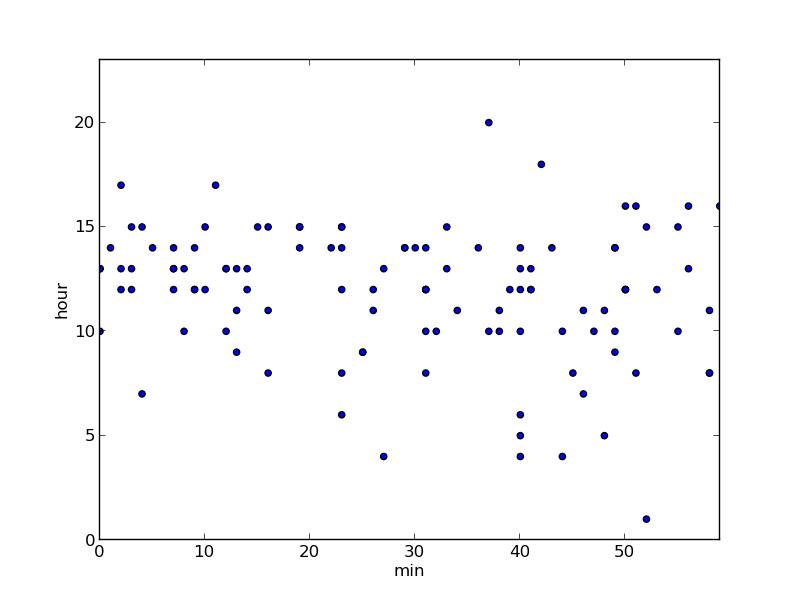}}
\label{fig:subfig2}
{\includegraphics[scale=.28]{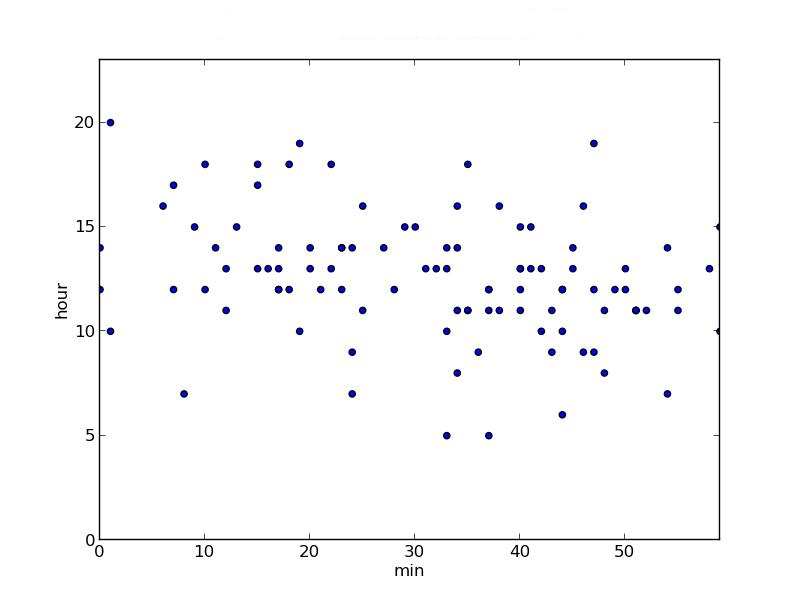}}
\label{fig:subfig3}
\caption[Average tweet time for bot and non-bot accounts]{Average Tweet Time for bot accounts (left) and non-bot accounts (right) using their data only from the events. (All bots and all non-bots from all events, some data points are overlapping).}
\label{fig:fake4}
\end{figure*}

We plotted graphs between inter-tweet delay mean and inter-tweet delay standard deviation for bot and non-bot accounts (see Figure \ref{fig:fake3}) using their data only from the events. Both bot and non-bot accounts have similar plots. To further analyze their tweeting pattern we also created average Tweet time plots (Figure \ref{fig:fake4}). Again we observed similar behavior between bot and non-bot accounts. 


\newpage
\section{Features}

Chu et. al. in their paper \cite{Chu:2010}, proposed a 4 part classification system to determine if a user is a bot or not. Chu et. al used a number of Twitter based features in their experiment. Namely:
\begin{enumerate}
\item Tweet Count: Number of tweets posted by an account. Bots post many more tweets than humans.
\item Long term hibernation by some accounts(in bots): Bots generally show long period of no activity and some short period of heavy activity. Humans generally have constant activity.
\item Ratio of Followers vs Friends: Bots tend to have more friends than followers. Humans on the other hand have nearly equal number of friends and followers. 
\item Temporal tweeting patterns: Bots are more active during specific days of the week. 
\item Account Creation Date: Bots are created more ``recently" than non-bot accounts. 
\item Device used for Tweeting: Bots generally Tweet via Twitter API or some other programmable services. Humans generally Tweet from mobile phones, web browsers etc.
\item Presence of URL in tweets: Bots tend to include more URLs in their tweets. 
\item Time trends: Zhang et. al. in their paper argue that because of the mode of automation bots are limited by the constraints of the Twitter API and will have to space out their tweets. They argued that bots show a pattern in their tweet time. 
\end{enumerate}

All the above mentioned features can be categorized into two, user based features and temporal based features. Features like Time trends, Temporal tweeting pattern, long term hibernation are all temporal based features and all others can be called user based features. 

In case of high impact events, many features stop contributing in helping us differentiate between bots and non-bot accounts. Some of the features that are not observed in High Impact events are (most of them have been discussed in previous sections):
\begin{enumerate}
\item Long term hibernation by some accounts: During high impact events all accounts that participate are not in ``hibernation". Bots that are active are the ones that are either already listening for updates or in some scenario created specifically for that event (example, spam bot @\_bostonmarathon discussed in Section \ref{rumor} was created specifically to Tweet during the Boston marathon).
\item Temporal tweeting patterns: A High Impact Event can occur any day of the week. Bot or non-bot accounts will post updates about these events irrespective of day of the week. 
\item Presence of URL in tweets: In Section \ref{URLA} we found that 83\% of bot tweets and 68\% of non-bot tweets have URLs in them. As this difference is quite small it can be argued that presence of URLs in tweets is not a big differentiator. 
\item Time trends: In Section \ref{TTA} we applied all analysis done by Zhang et. al. in their paper and found out that Time Tweeting pattern of accounts active during high impact events are very similar.
\end{enumerate}

\section{Real Time Prediction During an Event}
In this section we apply machine learning techniques to create a model that can help decide if an account is a bot or not. For our analysis, we created two set of features. One including temporal based features and one excluding temporal based features (including only user based features). We propose that user based features are a better indicator. We applied Decision Trees (J48) machine learning model to the dataset using WEKA.\footnote{\url{http://www.cs.waikato.ac.nz/ml/weka/}} We created a balanced model with equal numbers of bot and non-bot accounts. We ran a 10-fold cross validation scheme in which a subset of the original dataset is used for learning and another subset is used for evaluation of the decision tree model. 

The result of the classifier can be seen in table \ref{table:datar}. In the table \textbf{F1} represents the feature-set that includes only the user based features. \textbf{F2} represents the feature-set that includes both user and temporal based features. 

\begin{table}[!h]
\centering
\begin{center}
\end{center}
\begin{tabular}{|p{2cm}|p{1.3cm}|p{1.3cm}|}
\hline
\bf{}&\bf{F2}&\bf{F1}\\
\hline
Accuracy &66.54 \%  & 85.10\%\\
\hline
TP Rate &0.665 & 0.851\\
\hline
FP Rate &0.335 & 0.149\\
\hline
Precision &0.716 & 0.852\\
\hline
Recall &0.665 &0.851\\
\hline
F-Measure& 0.645 &0.851\\
\hline
ROC &0.788 & 0.913\\
\hline
\end{tabular}
\caption[Classification results]{F1: User features based classification results, F2: User and Temporal based classification results. TP Rate, FP Rate, Precision, Recall, F-Measure, ROC are represented as weighted averages of the two classes, bot and non-bot}\label{table:datar}

\end{table}

In table \ref{table:datar}, accuracy of F1 feature-set is higher than that of the F2 feature set. Using the same, we conclude that using user based features, we are able to better differentiate between bots and non-bots.

We also computed the best knowledge gain features using WEKA. For the F1 set, the best features in order of knowledge gain were, Device used for Tweeting, Presence of URLs, Ratio of Followers vs. Friends, Account creation date, and Tweet count. These results were similar to those obtained by Chu et. al. \cite{Chu:2010}.

\section{Detailed Analysis of Few Bots.}\label{5bots}
To further analyze how these bots that are active during high impact events, function in normal conditions, we picked up 5 bots (1 each from each event) and monitored them for a time period of ~1 month (5 March 2014 to 9 April 2014). We took a daily snapshot of their followers and tweets. We also collected all their mentions and any response by these bots during this time period. The only thing we didn't capture were Direct Messages sent to these bots by other users and vice-versa because we simply didn't have permissions to do so. \\

\begin{table}[!h]
\begin{center}
    \begin{tabular}{|l|l|l|l|l|}
    \hline
    User ID    & Screen Name   & Tweets                                         & Following & Followers \\ \hline
    21287212   & FintechBot    & 22.2K                                          & 297       & 1,137     \\ \hline
    219913533  & DTNUSA        & 565K                                           & 83        & 3,108     \\ \hline
    591713080  & tipdoge       & 23.9K                                          & 1,333     & 5,622     \\ \hline
    1348277670 & Warframe\_BOT & 3,397                                          & 1         & 4,251     \\ \hline
    606204776  & BBCWeatherBot & 535 (Deletes many tweets) & 29        & 307       \\ \hline
    \end{tabular}
\caption[The details of the 5 bots studied in depth]{The details of the 5 bots (As of April 9, 2014 2359 hrs IST).}
\end{center}
\end{table}

\begin{table}[!h]
\small
\begin{center}
    \begin{tabular}{|l|l|l|l|}
    \hline
    Screen Name    & Profile Description                                                                                                                                             & Location Mentioned & External Website                    \\ \hline
    @FintechBot    & A little twitter bot that curates&England&adendavies.com \\& financial services tech news. && \\& Also some curating by @Aden\_76.                                                                &             &                       \\ \hline
    @DTNUSA        & Comprehensive Daily News on &Canada &\\& United States of America Today && \\& \~ © Copyright (c) DTN News && \\&Defense-Technology News  && \\&http://defense-technologynews.blogspot.ca/      &              &  \\ \hline
    @tipdoge       & available commands: balance, deposit,&&tipdoge.info \\& withdraw, tip on the way to moon                                                                                         & ~                  &                         \\ \hline
    @Warframe\_BOT & This bot retweets ONLY Warframe ?Alerts&& \\& from @WarframeAlerts  Sorry, I know& &\\& multiple RT bug. I'll fix it.                                                       & ~                  & ~                                   \\ \hline
    @BBCWeatherBot & Ask us about the weather! Start a tweet & Everywhere&bbc.co.uk/weather\\& @BBCWeatherbot and tell us where && \\&and when you want to know about. && \\&Trial service by @NixonMcInnes && \\&@BBCRDat \#BBCconnected &          &                    \\ \hline
    \end{tabular}
\caption[The profile description, location, and external website mentioned by these 5 bots.]{The details of the 5 bots Part 2 (As of April 9, 2014 2359 hrs IST).}
\end{center}
\end{table}


@Warframe\_BOT has changed itself sometime after the event in which it was active and before the time of data collection. It has changed itself from a news aggregator website to a Twitter bot that tweets about game updates. It deleted all its previous tweets. Such behavior is quite common in bots. 

 \begin{figure*}[!h]
\centering
{\includegraphics[scale=.4]{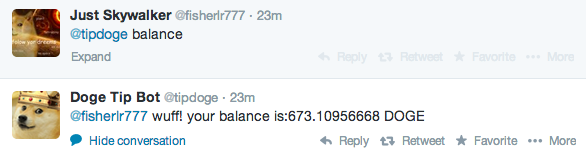}}
\label{fig:tip1}
{\includegraphics[scale=.4]{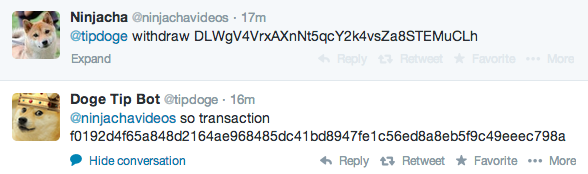}}
\label{fig:tip2}
\caption[Bot ``@tipdoge" interacting with users]{@tipdoge interacting with a user.}
\label{tippp}
\end{figure*}

\begin{figure*}[!h]
\centering
{\includegraphics[scale=.4]{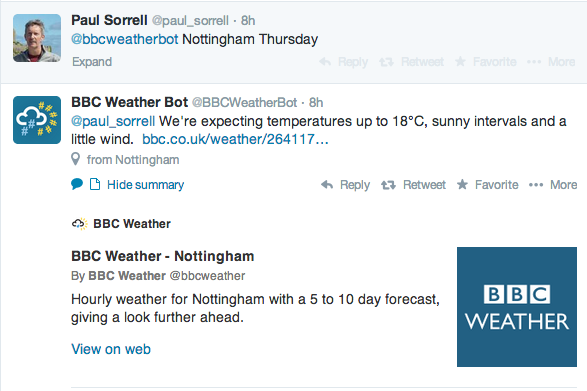}}
\label{fig:bbc1}
{\includegraphics[scale=.4]{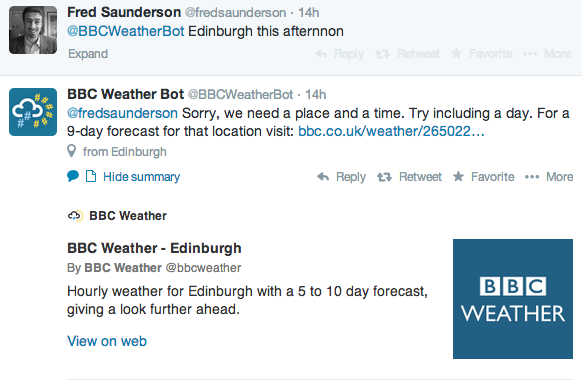}}
\label{fig:bbc12}
\caption[Bot ``@BBCWeatherBot" interacting with users]{@BBCWeatherBot interacting with a user.}
\label{fig:bbcccc}
\end{figure*}

We also recorded interactions that many users engaged with these 5 bots. These included Mentions, Retweets, and @replies. @FintechBot and @DTNUSA are bots that spread financial and general news respectively; users tend to Retweet their tweets and sometimes replied to their tweets. These bots never responded to these replies. This is a very frequently  observed pattern for these kind of bots. On the other hand  @BBCWeatherBo t and @tipdoge are another kind of bots which engage in user interactions. These bots are computer programs that require the user to input certain values in a particular format in the tweet to which they respond with some information. In case of @BBCWeatherBot it requires the user to tweet the name of the place and the bot returns the weather conditions in that place. @tipdoge similarly requires input and responds with the current status of or alteration to the user's dogecoin wallet.

\begin{figure}[!h]
\centering
\includegraphics[scale=.4]{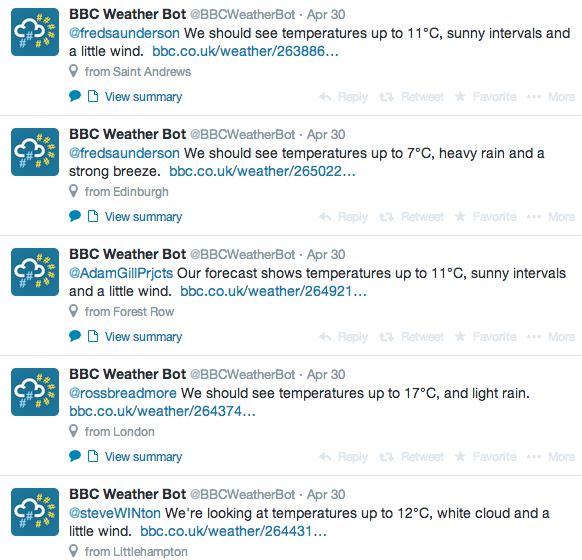}
\caption[Bot tweets having clear patterns and similarity]{Tweets by the @BBCWeatherBot showing clear patterns and text similarity.}
\label{fig:bbc2}
\end{figure}

We can also see that tweets by these bots have a pattern. They are highly repetitive in nature and also have very high Tweet similarity. This is classic bot behavior as they are nothing but programs that output Tweet texts and hence have a very limited output range.

\section{Growth and Changes Observed in Bots (2013 vs 2011)}
In order to determine how bots participate during high impact events have changed with years we compared bot activity during impactful events in 2013 with their activity in same kind of events from 2011. We took the data set for 2011 events from \cite{Gupta:2012:CRT:2185354.2185356}. 

\begin{table}[!h]
\begin{center}
    \begin{tabular}{|l|l|l|l|}
    \hline
    Event          & Hashtag/Trending Topic & Tweets  & Details\\ \hline
    Virginia Earthquake      & \#earthquake, Earthquake SF      & 277,604         & Magnitude 5.8 earthquake \\ \hline
    Indiana Fair Tragedy & Indiana State Fair      & 49,924        & Stage accident at Fair       \\ \hline
    Hurricane Irene           & Hurricane Irene       & 90,237        & Caused 55 deaths      \\ \hline
    Libya Crisis          & Libya, tripoli      & 389,506        & Rebel against Qaddafi     \\ \hline
    Mumbai Blast    & \#mumbaiblast,\#needhelp       & 32,156         & 3 bomb blasts      \\ \hline
    UK Riots        & \#ukriots, \#londonriots    & 542,685         & Riots in United Kingdom \\ \hline
    US Downgrading   & S\&P, AAA to AA       & 148,047         & Debt crisis in US   \\ \hline
    \end{tabular}
\caption{2011 crisis dataset.}
\label{table:2100c}
\end{center}
\end{table}

The data set comprised of data collected for the events mentioned in table \ref{table:2100c}. We used the same methodology to label accounts as bots and non-bots as used in the methodology section.

\begin{table}[!h]
\begin{center}
    \begin{tabular}{|l|l|l|}
    \hline
    ~               & Bots & Non-Bots\\ \hline
    All 2011 Events & 183     & 95        \\ \hline
    \end{tabular}
\caption[Annotation result on 2011 crisis dataset]{Bots in 2011 Crisis Dataset. The number of accounts identified as bots and non-bots.}
\end{center}
\end{table}

Please note we do not state that the analysis and the methodology stated in this section is correct because it has been 3 years since the data used in this section was collected. We do not say that the accounts that we labeled as bots or non-bots in 2014 behaved similarly in 2011. A lot has changed in 3 years; user tweeting pattern, Twitter API restrictions, etc. We were unable to collect all the tweets by these accounts from 2011. These accounts were annotated using their timeline in 2014.

We only compared the tweet sources used by these accounts. 

\begin{table}[!h]
\begin{center}
    \begin{tabular}{|l|l|l|}
    \hline
    Rank & 2011 Bot Sources & 2011 Non Bots Sources  \\ \hline
    1    & twitterfeed        & web                       \\ \hline
    2    & TweetDeck          & Twitter for iPhone        \\ \hline
    3    & web                & Twitter for BlackBerry    \\ \hline
    4    & Twitter for iPhone & Twitter for Android       \\ \hline
    5    & PageBase.Net       & Tweet Button              \\ \hline
    6    & Echofon            & Visibli                   \\ \hline
    7    & Tweet Button       & Twitter for iPad          \\ \hline
    8    & Resonancers        & UberSocial for BlackBerry \\ \hline
    9    & Mobile Web         & HootSuite                 \\ \hline
    10   & Butting In         & Mobile Web                \\ \hline
    \end{tabular}
\caption{Top tweet sources in 2011 crisis dataset by bots and nonbots.}
\label{table:2011src}
\end{center}
\end{table}

We can see in table \ref{table:2011src} that bots in 2011 used more Twitter Apps to post data to Twitter; this is in contrast to the use of many automation web apps like IFTTT and dlvr.it in 2013 (Table \ref{table:bostonsource}).

\chapter{Discussion, Limitations and Future Work}\label{chapter:discuss}
In this study, we have presented an analysis of automated activity during High Impact activity on Twitter. First, we begin by defining high impact events and highlighting how Twitter is used as a news medium by its users for various updates during high impact events. We then discussed use of bots on Twitter and some Twitter API rules that govern posts on the online social network; these rules are very important as they help regulate bot activity on Twitter. 

In the next part of the study we discussed our methodology and criteria of event selection based on their political and economical impacts and importance. We collected the data for each of the event through the Twitter API. We also discussed our annotation scheme in which we got about 1,000 Twitter accounts annotated in two categories as ``bots" and ``non-bots". As a result of annotation we got 377 bot accounts and 115 non-bot accounts. We then used the Twitter API to collect more data associated to these annotated accounts like their Timeline, followers etc. 

In Chapter 5 we present our analysis of automated activity during high impact events on Twitter. We begin by discussing how to create bots, and how this process is constrained by various Twitter API rules. We then moved on to discuss some basic characteristics of these bots and non-bots on Twitter. We presented some insights on number of Friends and Followers of these annotated accounts, and also, the most common keywords that are present in their profile description. We then discussed the network formed by bots and their followers and friends during the Boston marathon bombings and how these bots ``broker" information from some trusted accounts to the masses. We also showed that bots actually did not propagate rumors and even if they do they pick them very late, this can be attributed to the fact that they mainly use trusted sources for the information that they propagate in the network. We also analyzed their URLs, Tweet sources and Tweet times. We also discussed in detail user and temporal based features that can be used to create a classifier to decide if a given account is a bot or not, we used WEKA to create this classifier which had an accuracy of 85.10 \% using user based features and not temporal based features. We also analyzed 5 bots active during high impact events in detail and show how they actually work. We also compared bot activity between 2011 and 2013. 

We must also highlight that a possible limitation of our work can be the fact that annotations were done after a few months of the event. Bot behavior could have changed or some other conditions altered during this time interval. Also, we cannot claim that data collected from Twitter API endpoints, which gives limited access to the actual data, can be said to be representative of the actual data. 

In future, we would also like to analyze more events to strengthen our analysis and assertions. We would like to verify these results for other high impact events. We would also like to investigate more on the impact of automated activity in propagation of fake and malicious content on Twitter. We will also like to develop a tool to differentiate between bot and nonbot accounts in real time on Twitter.

\bibliographystyle{acm}
\bibliography{icwsm}

\chapter{Appendix}\label{chapter:appendix} 

\section{Tweet Sources for Other Events}
\begin{table}[!ht]
\begin{center}
    \begin{tabular}{|l|l|l|l|l|}
    \hline
    Rank & Icestorm Bots Source & Count & Icestorm Non-bots Source & Count \\ \hline
    1    & web                  & 9858  & Twitter for iPhone       & 959   \\ \hline
    2    & PageBase.Net         & 5676  & web                      & 587   \\ \hline
    3    & RoundTeam            & 2187  & Twitter for Android      & 65    \\ \hline
    4    & twitterfeed          & 2149  & HootSuite                & 13    \\ \hline
    5    & Sam Tweet            & 2034  & TweetDeck                & 12    \\ \hline
    6    & SweeterTweet         & 1511  & Twitter for BlackBerry   & 8     \\ \hline
    7    & Buffer               & 1488  & Twitter for iPad         & 7     \\ \hline
    8    & RoundTeam            & 1218  & Mobile Web               & 4     \\ \hline
    9    & dlvr.it              & 1058  & Twittascope              & 3     \\ \hline
    10   & TweetWithTag         & 775   & Instagram                & 3     \\ \hline
    \end{tabular}
\caption{Top Tweet Sources for Icestorm dataset.}
\end{center}
\end{table}

\begin{table}[!ht]
\begin{center}
    \begin{tabular}{|l|l|l|l|l|}
    \hline
    Rank & Navy Yard Bots Sources & Count & Navy Yard Non-Bot Sources & Count \\ \hline
    1    & twitterfeed            & 2429  & Twitter for iPhone        & 385   \\ \hline
    2    & web                    & 806   & RoundTeam                 & 384   \\ \hline
    3    & IFTTT                  & 662   & web                       & 203   \\ \hline
    4    & dlvr.it                & 563   & twitterfeed               & 176   \\ \hline
    5    & RoundTeam              & 453   & Twitter for Android       & 58    \\ \hline
    6    & TweetDeck              & 373   & Twitter for iPad          & 55    \\ \hline
    7    & TweetAutoPilot         & 297   & HootSuite                 & 51    \\ \hline
    8    & Crime News Updates     & 261   & Ask.fm                    & 31    \\ \hline
    9    & Tweet Old Post         & 259   & Vine - Make a Scene       & 13    \\ \hline
    10   & Trendsmap Alerting     & 180   & Tweet Button              & 10    \\ \hline
    \end{tabular}
\caption{Top Tweet Sources for Navy Yard dataset.}
\end{center}
\end{table}

\begin{table}[!ht]
\begin{center}
    \begin{tabular}{|l|l|l|l|l|}
    \hline
    Rank & Oklahoma Bots Sources & Count & Oklahoma Non Bots Sources & Count \\ \hline
    1    & twitterfeed           & 728   & web                       & 618   \\ \hline
    2    & web                   & 489   & Twitter for iPhone        & 440   \\ \hline
    3    & dlvr.it               & 410   & Twitter for Android       & 374   \\ \hline
    4    & RoundTeam             & 253   & TweetDeck                 & 208   \\ \hline
    5    & WordPress.com         & 120   & Twitter for iPad          & 199   \\ \hline
    6    & I-44 Update           & 119   & HootSuite                 & 86    \\ \hline
    7    & IFTTT                 & 117   & The Huffington Post       & 74    \\ \hline
    8    & SimpleWeatherAlert    & 116   & Twitter for BlackBerry    & 46    \\ \hline
    9    & twicca                & 92    & Wunderground              & 43    \\ \hline
    10   & Crime News Updates    & 87    & Tweet Button              & 42    \\ \hline
    \end{tabular}
\caption{Top Tweet Sources for Oklahoma dataset.}
\end{center}
\end{table}

\begin{table}[!ht]
\begin{center}
    \begin{tabular}{|l|l|l|l|l|}
    \hline
    Rank & Phailin Bots Source   & Count & Phailin Non Bots Sources   & Count \\ \hline
    1    & twitterfeed          & 658   & Twitter for Android       & 339   \\ \hline
    2    & dlvr.it              & 417   & web                       & 305   \\ \hline
    3    & SCOOP Hot News India & 407   & TweetDeck                 & 293   \\ \hline
    4    & Retweet-phailin      & 288   & Twitter for iPhone        & 147   \\ \hline
    5    & dlvr.it              & 214   & Twitter for BlackBerry    & 108   \\ \hline
    6    & web                  & 207   & Twitter for Windows Phone & 84    \\ \hline
    7    & Tweet Old Post       & 131   & Twitter for iPad          & 65    \\ \hline
    8    & RoundTeam            & 122   & Nimbuzz Mobile            & 59    \\ \hline
    9    & TrendTweeter         & 88    & Mobile Web (M2)           & 48    \\ \hline
    10   & WordPress.com        & 75    & HootSuite                 & 43    \\ \hline
    \end{tabular}
\caption{Top Tweet Sources for Phailin dataset.}
\end{center}
\end{table}

\newpage
\section{IFTTT Recipes }

\begin{figure}[!h]
\centering
\includegraphics[scale=.3]{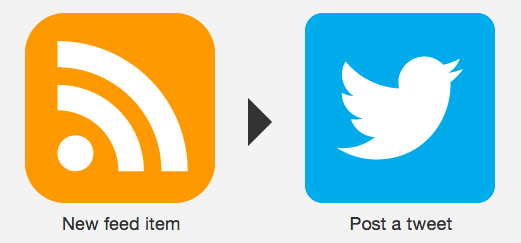}
\caption{An IFTTT Recipe to post a RSS Feed item as a Tweet.}
\label{fig:ifttt1}
\end{figure}

\begin{figure}[!h]
\centering
\includegraphics[scale=.3]{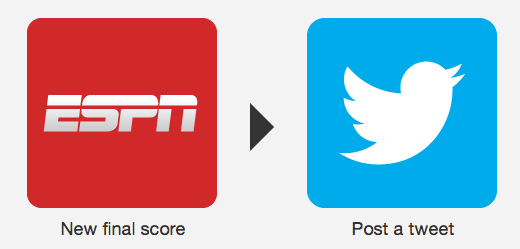}
\caption{An IFTTT Recipe to automatically post a new updated scores as a Tweet.}
\label{fig:ifttt1}
\end{figure}

\begin{figure}[!h]
\centering
\includegraphics[scale=.3]{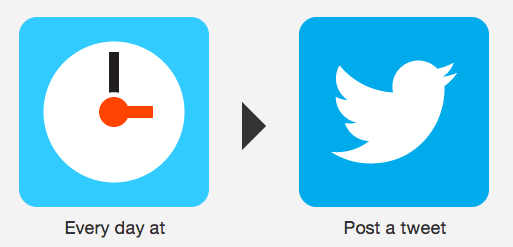}
\caption{An IFTTT Recipe to schedule tweets.}
\label{fig:ifttt1}
\end{figure}

\begin{figure}[!h]
\centering
\includegraphics[scale=.3]{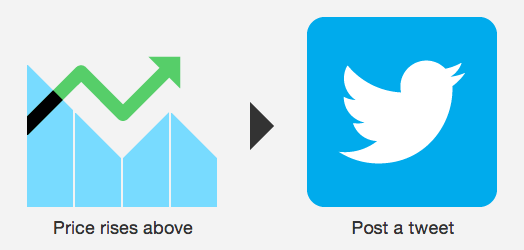}
\caption{An IFTTT Recipe to post tweets when a share price goes above a particular price.}
\label{fig:ifttt1}
\end{figure}

\end{document}